\documentclass[journal]{IEEEtran}%[times, 10pt,twocolumn]{article}
\usepackage{times,psfrag,subfigure,amssymb,amsmath,shortvrb,graphicx,alltt,comment}
\providecommand{\OO}[1]{\mathop{\mathrm{O}}\bigl(#1\bigr)}
\begin{document}
\def\reffig#1{Fig.\,\ref{#1}}
\def\reffeq#1{(\,\ref{#1})}
\def\reftable#1{Table.\,\ref{#1}}
\title{Lossy Bulk Synchronous Parallel Processing Model for Very Large Scale Grids}
\author{Elankovan Sundararajan, Aaron Harwood, Kotagiri Ramamohanarao
\thanks{E. Sundararajan, A. Harwood and Kotagiri Ramamohanarao
are with the Department of Computer Science and Software
Engineering, The University of Melbourne, ICT Building, 111 Barry
Street, Carlton 3053, Victoria Australia. Email: \{esund,aharwood,
rao\}@csse.unimelb.edu.au.}}

\markboth{IEEE TRANSACTION ON PARALLEL AND DISTRIBUTED SYSTEMS
,~Vol.~-,No..~-,~Month~-}{Shell \MakeLowercase{\textit{et al.}}:
UDP-based Parallel Computing on Very Large Scale Grids}

\maketitle

\begin{abstract}
The performance of a parallel algorithm in a very large scale grid
is significantly influenced by the underlying Internet protocols
and inter-connectivity. Many grid programming platforms use TCP
due to its reliability, usually with some optimizations to reduce
its costs. However, TCP does not perform well in a high bandwidth
and high delay network environment. On the other hand, UDP is the
fastest protocol available because it omits connection setup
process, acknowledgments and retransmissions sacrificing reliable
transfer. Many new bulk data transfer schemes using UDP for data
transmission such as RBUDP, Tsunami, and SABUL have been
introduced and shown to have better performance compared to TCP.
In this paper, we consider the use of UDP and examine the
relationship between packet loss and speedup with respect to the
number of grid nodes. Our measurement suggests that packet loss
rates between $5\%$-$15\%$ on average are not uncommon between
PlanetLab nodes that are widely distributed over the Internet. We
show that transmitting multiple copies of same packet produces
higher speedup. We show the minimum number of packet duplication
required to maximize the possible speedup for a given number of
nodes using a BSP based model. Our work demonstrates that by using
an appropriate number of packet copies, we can increase
performance of parallel program.
\end{abstract}


\section{Introduction}

\PARstart{P}{arallel} computing has become increasingly popular
due to widespread availability of cost effective computational
resources, such as commodity SMPs, PCs and high-performance
cluster platforms. The size of individual clusters has also
continued to grow as evidenced by data collected from the top 500
supercomputers~\cite{bouteiller03coordinated}. This has benefited
large-scale application research that was once accessible only to
a relatively small number of researchers.

However, as the size of computational grids continues to grow, to
become very large scale grids (VLSG), the number of wide area
network (WAN) connections between islands of clusters and other
high performance computing (HPC) centers grows quadratically to
the number of nodes. These WAN connections put limits on the
granularity of parallel applications that could otherwise benefit
from the available computing power, i.e. computation to
communication ratio needs to be significantly large in order for
the communication complexity to not dominate the run-time.
Embarrassingly parallel, data parallel and parametric problems
that do not require significant message passing can be efficiently
parallelized but problems that require significant communication
present challenges. It is important to understand how these
problems can be efficiently parallelized. The approach we consider
in this paper is to understand the effect of the WAN connections
by examining the relationship of network bandwidth, delay, and
loss of packets with speedup.
%An orthogonal approach which we
%consider in other work is to adapt the granularity of the
%solutions, e.g. scaling the problem size.

Transmission Control Protocol (TCP)~\cite{Postel81} and User
Datagram Protocol (UDP)~\cite{Postel80} are currently the
predominant protocols used for end-to-end communication. TCP
provides useful services such as connection-oriented, streaming,
full-duplex, reliable, and end-to-end semantic to its applications. These
services provide reliability at a cost, causing delay in
transmission. Some of these services can be sacrificed depending
on the types of applications executed over WAN. Typical grid programming
platforms use TCP or TCP with some optimization, but TCP does
not perform well in a high bandwidth and high delay network
environment~\cite{Irwin92,Jacobson92}.

As network bandwidth increases rapidly together with the advent of
new routing/switching technology like Multi-protocol Label
Switching (MPLS), the load on end systems and the data transfer
protocols are becoming bottlenecks in many cases~\cite{Yuhong04}.
This indicates performance of parallel programs (mainly
constrained by communication phase) on WAN is no longer hardware
constrained. Thus we emphasize our study on the bottleneck caused
by the data transfer protocol with assumption that end systems
with manageable load are used in computing on WAN.

TCPs congestion control algorithm (exponential back-off) causes
packet transfer throughput to collapse even when the bandwidth is
still plentiful. It is important to realize that packet losses do
not necessarily happen just because of network congestion. It is
well known that TCP was originally designed for reliable data
communication on low bandwidth and high error rate
networks~\cite{Postel81}. A packet is retransmitted when it gets
corrupted or lost. The reliability provided by TCP reduces network
throughput, increases average delay and worsens delay
jitter~\cite{Feng00,Liu02}. While reliable transmission is quite
often critical for the proper execution of a parallel process, use
of TCP is not the only means of attaining reliability and it is
not clear whether TCP algorithms in general are the correct
approach with respect to communication patterns of parallel
processes. It is generally accepted that TCP is not suitable for
delay constrained applications that emphasize performance issues.

UDP on the other hand tends to be the fastest protocol because it
omits connection setup process, acknowledgments, and
retransmission. Each packet sent is independent of all other
packets. However, unlike TCP, packet losses can occur and a
mechanism has to be provided to take necessary measure to detect
and assure successful delivery of packets~\cite{Postel80}.

Many high performance data transfer protocols have been developed
using UDP recently. The Tsunami\cite{Mark} reliable file transfer
protocol was developed for high-bandwidth dedicated research
networks that never experience significant congestion. It contains
two user-space applications (a client and a server) and uses UDP
for data transfer and TCP for sending controls information (such
as retransmission request, restart request, error report and
completion report). This protocol uses inter-packet delay as means
of flow control as opposed to sliding-window mechanism in TCP.
Another protocol known as Reliable Blast UDP (RBUDP)\cite{He02} is
an aggressive bulk data transfer scheme. It is intended for
extremely high bandwidth, dedicated or Quality-of-Service enabled
networks such as optically switched networks. This scheme sends
the entire payload at a user-specified sending rate using UDP
datagrams and TCP is used to send signal indicating the end of
transmission from sender and acknowledgment from receiver
consisting bitmap tally of the received packets. This work also
demonstrates that the load factor of receiving node contributes to
packet loss rate. Simple Available Bandwidth Utilization Library
(SABUL) \cite{Yuhong04} is another high performance data transfer
protocol for data intensive application over high bandwidth
network. This reliable and lightweight application protocol uses
UDP for data transfer and TCP for feedback control messages. It
also uses rate based congestion control mechanism that tunes the
inter-packet time as in Tsunami. All this work shows that the use
of UDP with some reliability can enhance performance for WAN based
application that require immense data movements. Thus, we
investigate the performance of UDP for parallel computing over
WAN.

Our work considers the possibility of achieving good parallel
processing speedup using UDP on a WAN with some reliability via
acknowledgment packet. We investigate the effect of packet loss on
speedup and present a variant of the Bulk Synchronous Parallel
(BSP) processing model that considers packet loss as a fundamental
parameter. This is because we believe packet loss is the main
contributor in performance of UDP based transmission since it is
an unreliable protocol. To counter the unreliability of UDP we
introduce an acknowledgment packet from the receiver. Thus the
packet sending nodes involved in the communication phase knows if
a packet is lost (i.e. a packet is not received by a receiving
node).

\subsection{UDP measurements on PlanetLab}

\begin{figure*}
\centering
\includegraphics[height=2.5in,width=4.3in]{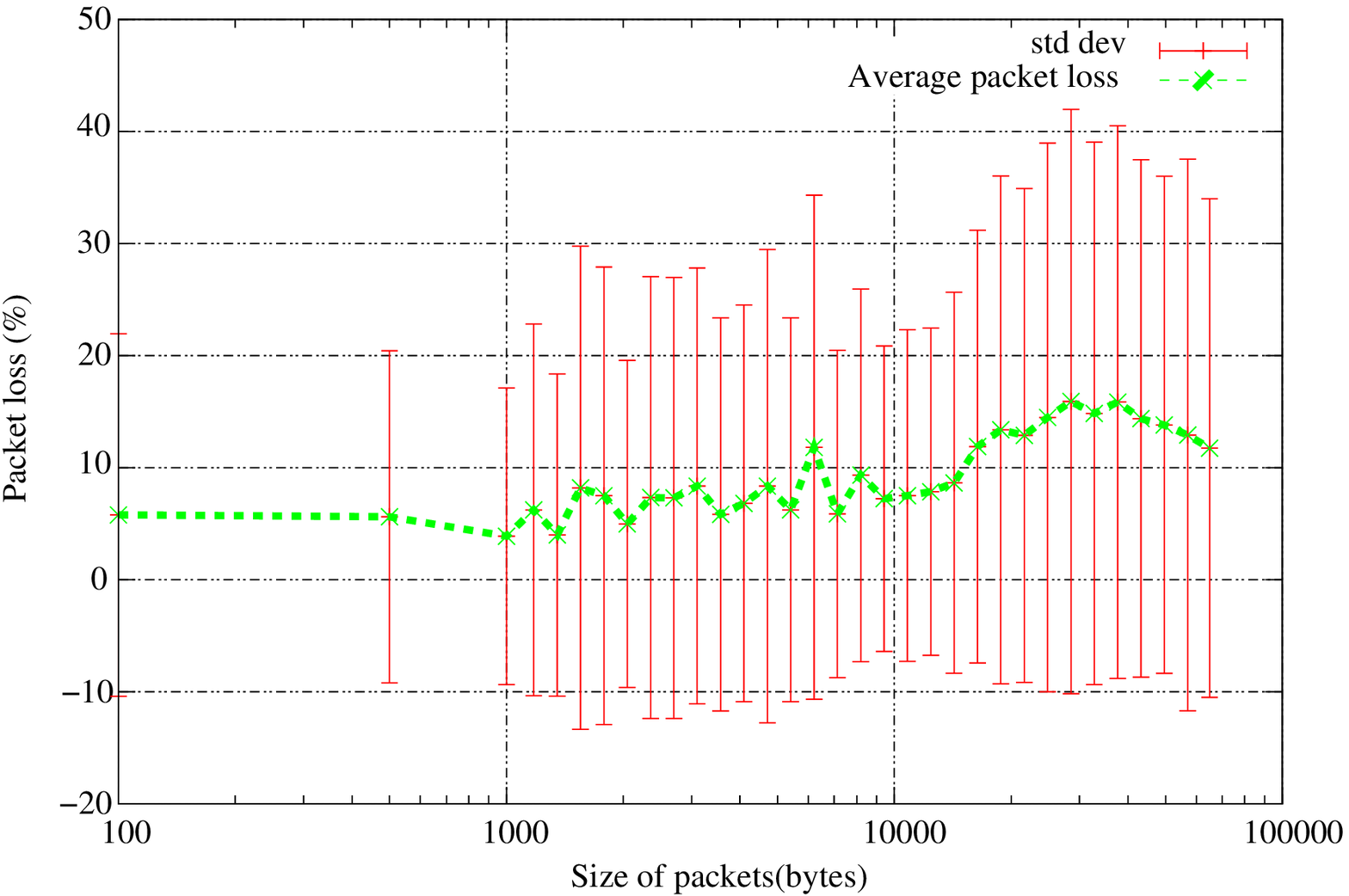}
\caption{\label{packloss}Average packet loss between pairs of
nodes for UDP based transmission within PlanetLab.}
\end{figure*}

\begin{figure*}
\centering
\includegraphics[height=2.5in,width=4.3in]{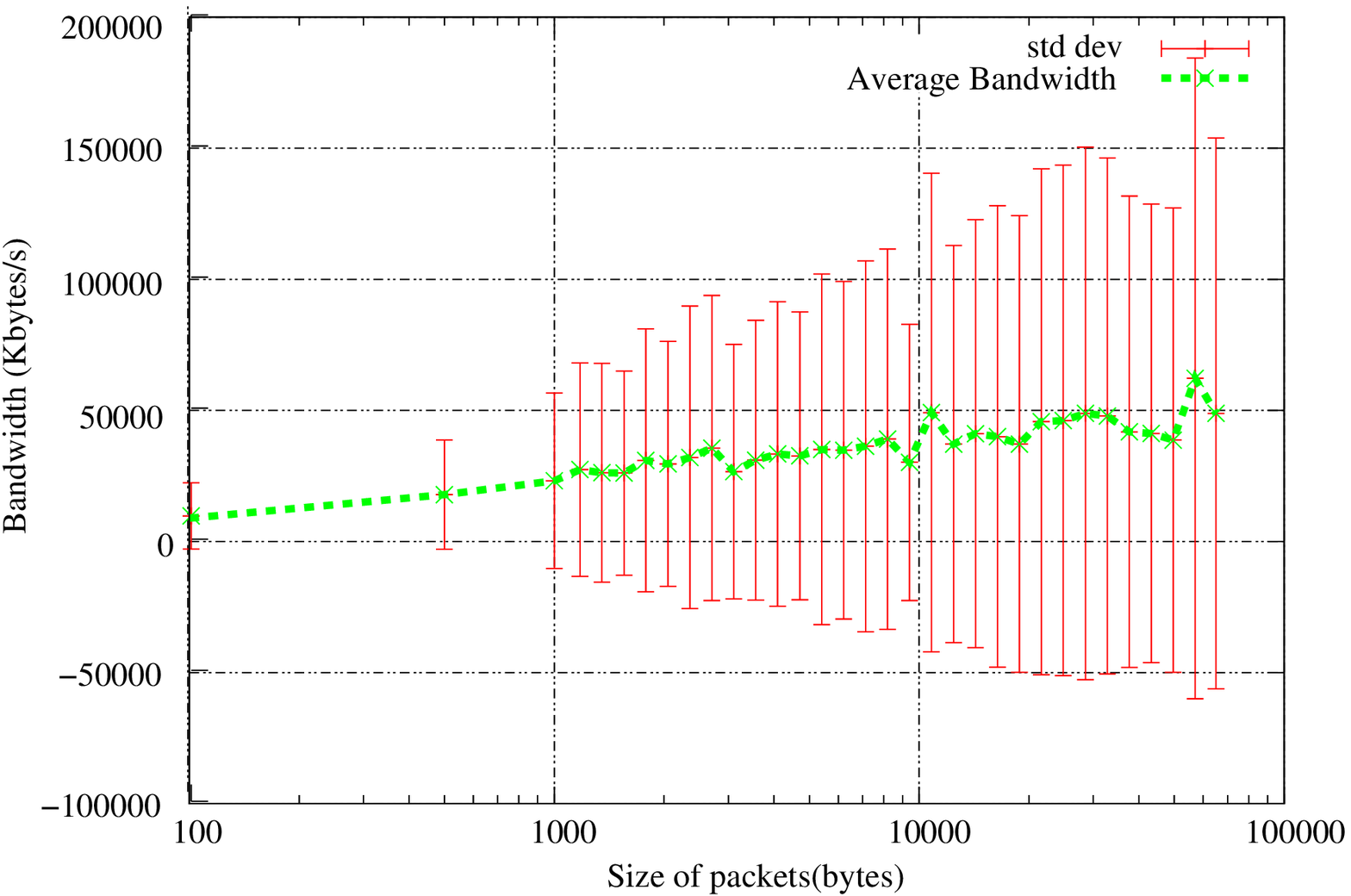}
\caption{\label{bw}Average bandwidth between pairs of nodes for
UDP based transmission within PlanetLab.}
\end{figure*}

\begin{figure*}
\centering
\includegraphics[height=2.5in,width=4.3in]{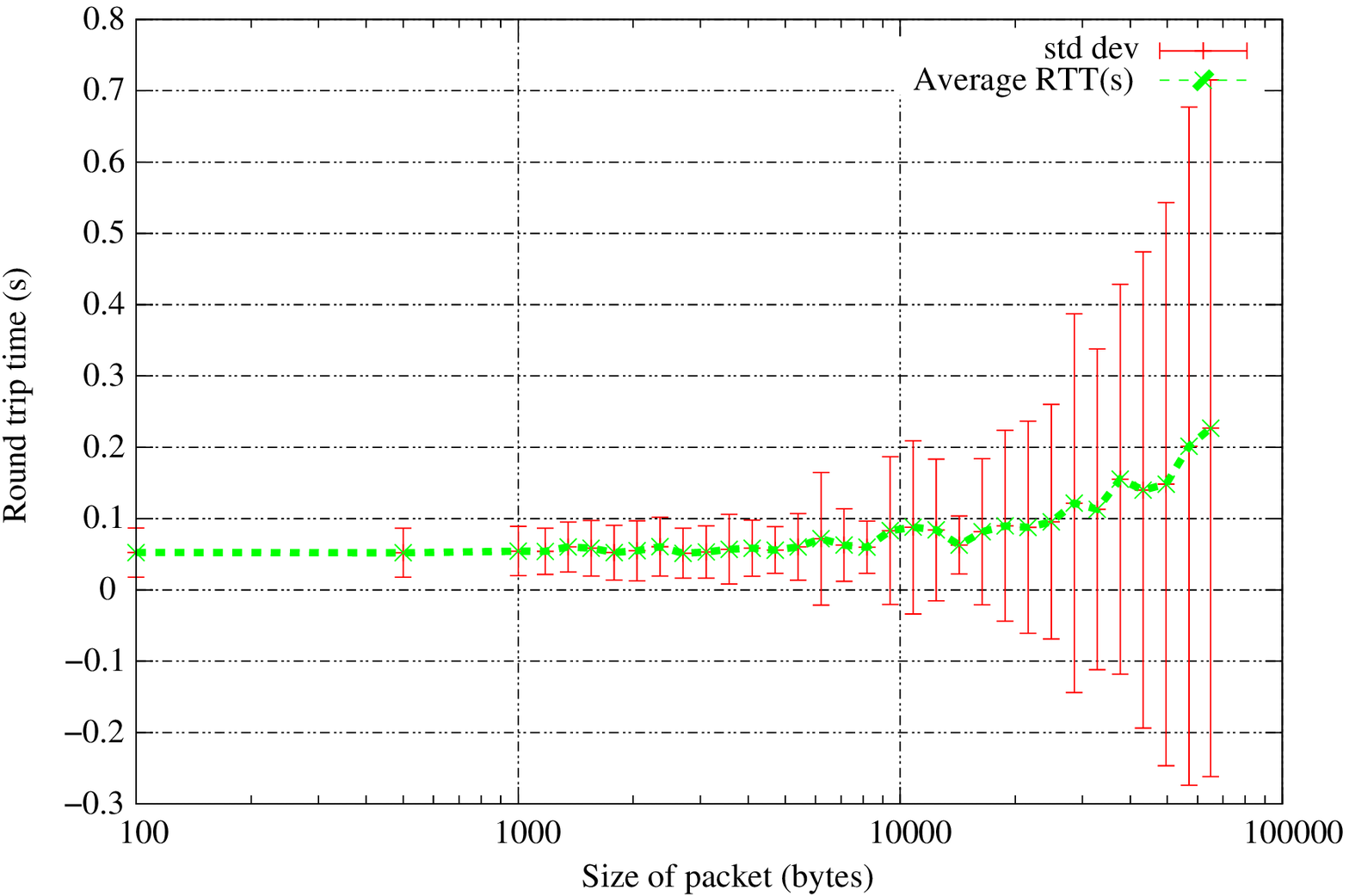}
\caption{\label{rtt}Average round trip time between pairs of nodes
for UDP based transmission within PlanetLab.}
\end{figure*}

To obtain a realistic view of packet loss, bandwidth and
round-trip-time on a very large scale grid, we measured UDP
behavior for different sizes of packets, between randomly selected
nodes from the top level domain ending with ``.edu" within
PlanetLab~\cite{PlanetLab}. We used utility programs and scripts
that we developed to select pairs of nodes randomly from almost
$160$ ``.edu" nodes. These nodes are then used to measure end to
end packet loss, round-trip-time and bandwidth using UDP. $100$
pairs of nodes were used in the experiment and each pairs were run
one at a time. Average packet losses between $5\%$-$15\%$ are
registered on this platform, plotted in~\reffig{packloss}. It is
also interesting to note that the percentage of packet loss are
independent of packet size for up to $10$ k/bytes, with loss of
less than $10\%$ and increases slightly to about $15\%$ for larger
packet sizes. There are cases when packet losses exceeds $15\%$,
this is probably due to high load and physical links on end
systems. Bandwidth and round-trip-time was also measured for UDP
on PlanetLab as depicted in~\reffig{bw} and~\reffig{rtt}
respectively. We observed that bandwidth of $30$Mbytes per second
to $50$Mbytes per second on average can be achieved using UDP.
Round-trip-time between $0.05$s and $0.1$s on average are recorded
for packet sizes of up to $25$Kbytes. Using these information we
analyzed the best speedup that can be achieved for differing
packet loss probability. The extent to which these measurements
are indicative of the Internet in general is to be further
investigated, though it is reasonable to suggest that a large
scale, shared grid system will exhibit similar behavior.

In section 2, we introduce stochastic models to analyze the impact
of packet loss on speedup; section 3 explains how we derived the
number of packet copies to use for maximizing the speedup; section
4 analyzes speedup of parallel computation when only lost packets
are re-transmitted; section 5 discuss some related work on
traditional systems and on WAN systems; and in section 6 we
summarize our conclusions and future work.

\section{The approach}

This section introduces a couple of approach used to evaluate
performance of parallel algorithms that use UDP protocol for
communication purposes. We begin with a conceptual approach and
later to a more realistic BSP~\cite{valiant90} based model that
reflects the effect of packet loss. In our analysis, we used the
notion of sending multiple copies of the same packet such that the
probability of packet loss approaches zero.

\begin{figure}
\centering
\psfrag{(1-p)}{$(1-p)$}\psfrag{(1-p)^2}{$(1-p)^2$}\psfrag{(1-p)p}{$(1-p)p$}
\psfrag{p}{$p$}\psfrag{Data sent and received}{Data sent and
received}\psfrag{Ack sent and received}{Ack sent and received}
\psfrag{Data sent but not received}{Data sent but not received}
\psfrag{Ack sent but not received}{Ack sent but not received}
\psfrag{Sender}{Sender} \psfrag{Receiver}{Receiver}\psfrag{?}{$?$}
\psfrag{Probability of}{Probability of}
\psfrag{transmission}{transmission} \psfrag{I}{i)}
\psfrag{II}{ii)} \psfrag{III}{iii)}
\includegraphics[height=2.8in,width=2.6in]{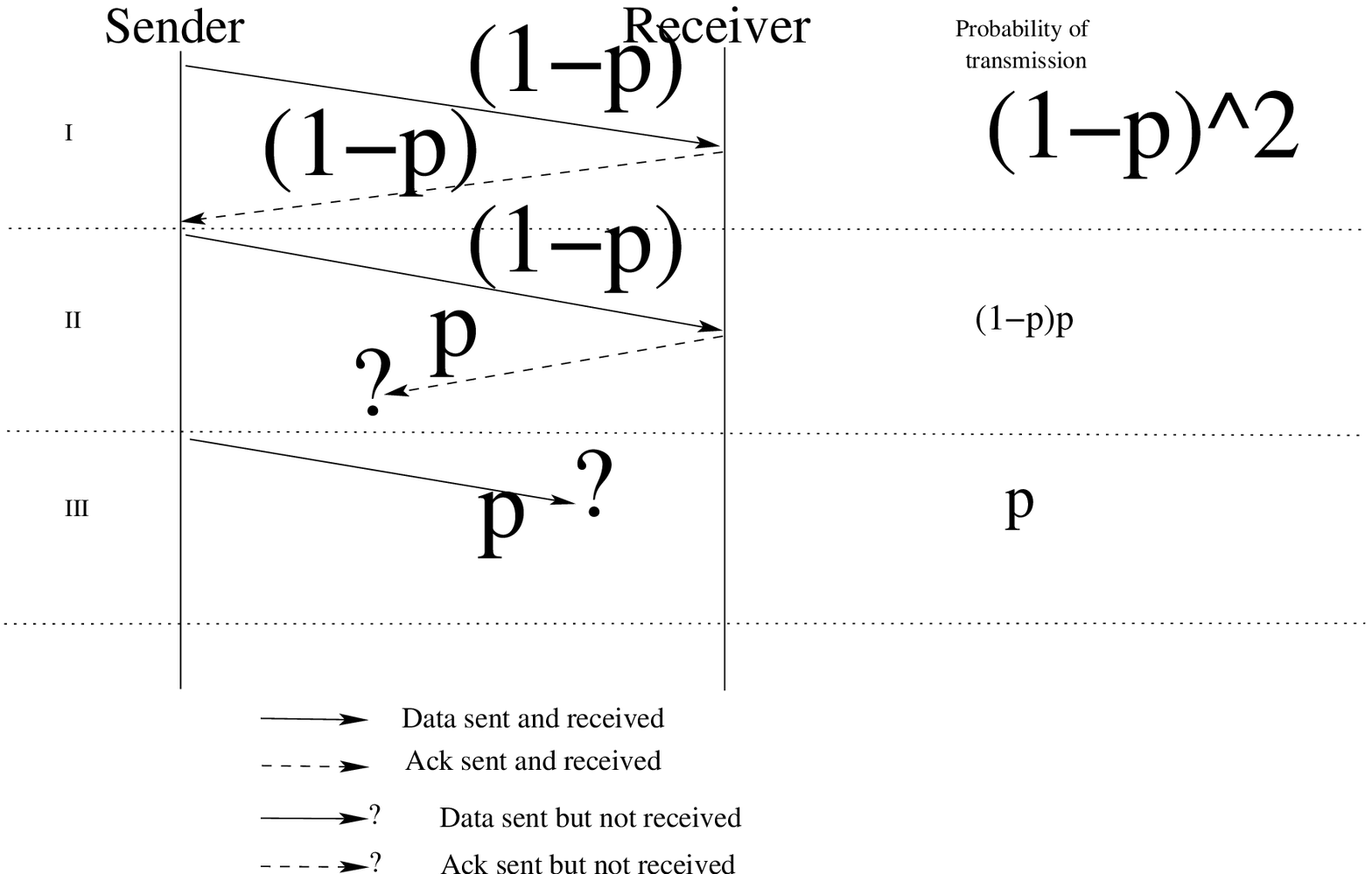}
\caption{\label{pack_loss_scen}Different packet loss scenario for
data packet and acknowledgment packet.}
\end{figure}

Consider sending a packet of data between $2$ nodes with a fixed
loss probability of $p$. We assume probability of data packet loss
and acknowledgment packet loss are identical. There are three
scenarios that could happen as depicted in
\reffig{pack_loss_scen}: \emph{i)} The data packet sent by sender
is successfully received by the receiving node, and an
acknowledgment packet sent by the receiver is successfully
received by the sender. This happens with a probability of
$(1-p)^2$.\emph{ ii)} The data packet sent by sender is
successfully received by the receiver but the acknowledgment
packet sent by the receiver is lost, with a probability of
$(1-p)p$. \emph{iii)} The data packet sent by the sender is lost,
with probability of $p$.

The probability of successful delivery is thus given by $(1-p)^2$.
Let $c(n)$ be the total number of packets transmitted for a
particular communication primitive involving $n$ processors. The
probability of successful delivery of $c(n)$ data packets and
successfully receiving the acknowledgment packets is
$p_s(n,p)=(1-p)^{2c(n)}$ and the probability of at least one
packet loss is $p_f(n,p)=1-p_s(n,p)$. Ideally, as $n
\longrightarrow \infty$, we have maximum capacity for speedup in
computing, however $p_f(n,p)\longrightarrow 1$ and thus the system
fails to operate. As such, we consider the best $n$ to use for a
particular communication complexity of $c(n)$.
\begin{figure}
\centering
\psfrag{w}{$w$}\psfrag{c(n)}{$c(n)$}\psfrag{1}{$1$}\psfrag{2}{$2$}\psfrag{r}{$r$}
\includegraphics[height=1.5in,width=2.4in]{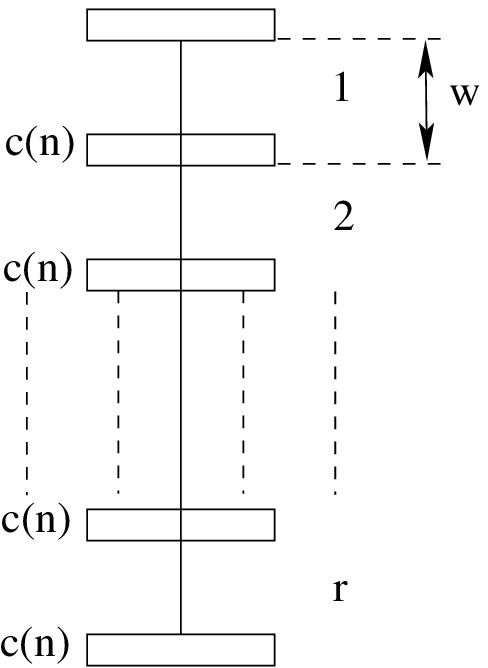}
\caption{\label{simpleCom}Computation, $w$ and communication,
$c(n)$ are performed $r$ rounds.}
\end{figure}
\subsection{The conceptual approach}
In this section we introduce a simplified conceptual notion that
communication between computing nodes are zero (an ideal
environment for parallel computing). This approach is similar to
that of PRAM model that provided the impetus for the existence of
better parallel computing models. Our approach here is not
suitable for practical purposes, but will be useful to help
understand the theoretical approach. The computation and
communication are performed for $r$ rounds, (see
\reffig{simpleCom}). The sending node only knows if a round has
failed (i.e. at least one packet is lost in the round). When a
round has failed, computation $w$ ($w$ is measured in seconds of
work on a processor) is performed again and $c(n)$ packets are
retransmitted. Here computation is performed again so as to
provide some penalty for the packet losses. The conceptual
approach is used to predict the best number of computing nodes to
use when communication is assumed to be negligible. Let $T(1)=wr$
represent the time taken to perform computation on a single node,
thus $T(n)=\frac{wr}{n}$ is the time taken to perform the same
computation on $n$ processors. Since $(1-p_s(n,p))^{i}p_s(n,p)$ is
the probability that $i$ attempts fail and the $(i+1)$-th attempt
succeeds, we have:
\begin{equation}\label{N_of_transm}
\hat{\rho}=\sum_{i=1}^{\infty}ip_f(n,p)^{i-1}p_s(n,p) =
\frac{1}{p_s(n,p)}.
\end{equation}
$\hat{\rho}$ gives the expected number of times all the packets
are transmitted. Therefore, expected time taken on $n$ processor
with packet loss probability $p$, $\hat{T}(n,p)$, is given by
~$\hat{T}(n,p)=w\hat{\rho} r=\frac{wr}{np_s(n,p)}$ to represent
the total time taken to complete the computation on $n$
processors. Hence, expected speedup of
$S_E=\frac{T(1)}{\hat{T}(n,p)}=np_s(n,p)$ can be achieved. Using
this expected speedup for different communication $c(n)$, we can
determine the optimal number of nodes, $n$, by solving
$\frac{\partial{S_E}}{\partial{n}}=0$ for $n$.

\begin{figure*}[htbp]
\centerline{\subfigure[\label{simp_comm_klogn}$c(n)=1$]{\includegraphics[height=2.1in,width=2.1in]{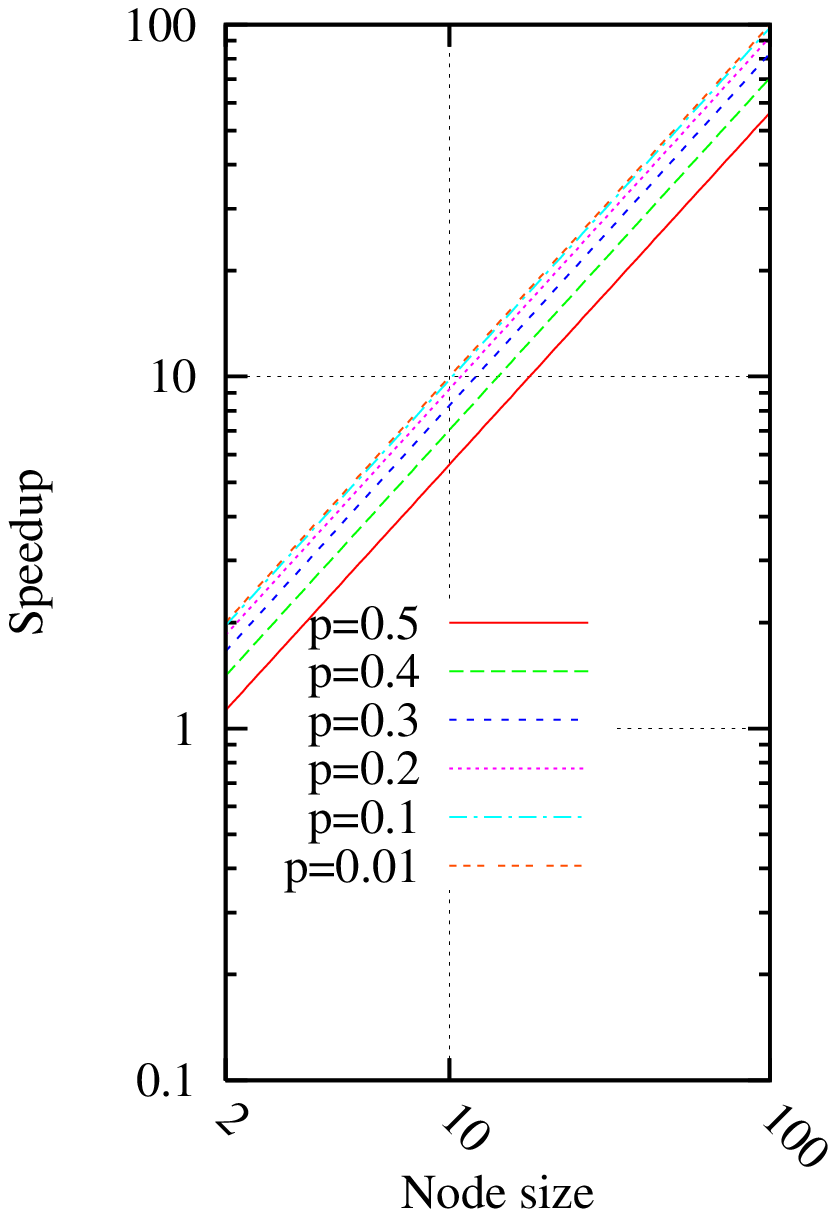}}
\subfigure[\label{simp_comm_klogn}$c(n)=log_2(n)$]{\includegraphics[height=2.1in,width=2.1in]{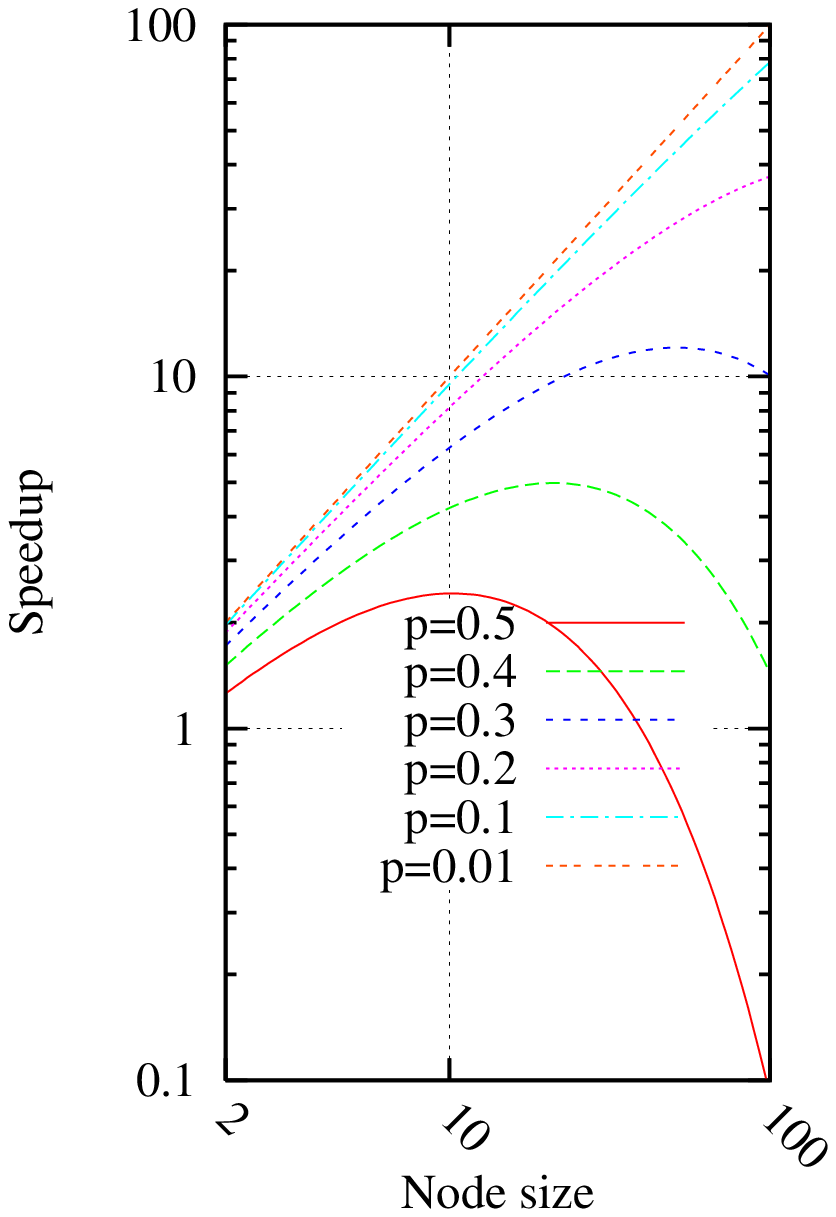}}
\subfigure[\label{simp_comm_klogn2}$c(n)=log_2^2(n)$]{\includegraphics[height=2.1in,width=2.1in]{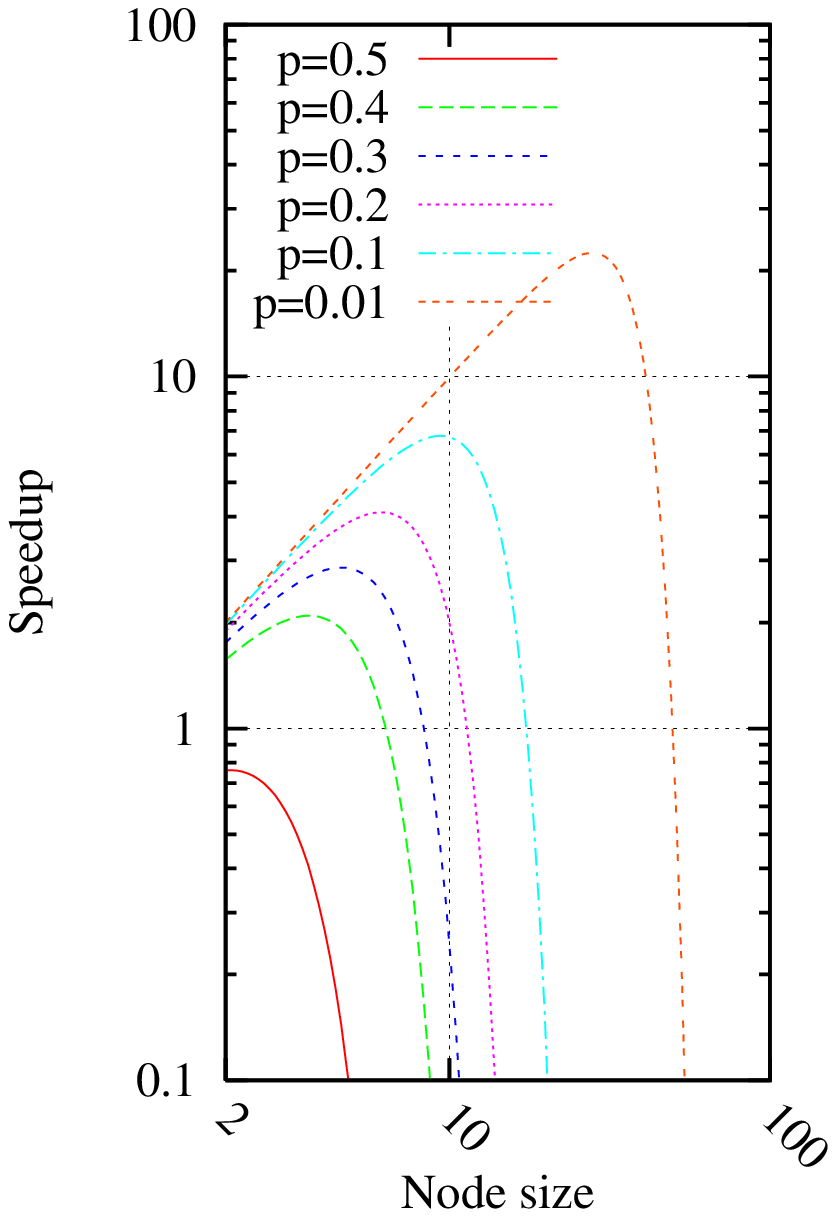}}}
\centerline{\subfigure[\label{simp_comm_kn}$c(n)=n$]{\includegraphics[height=2.1in,width=2.1in]{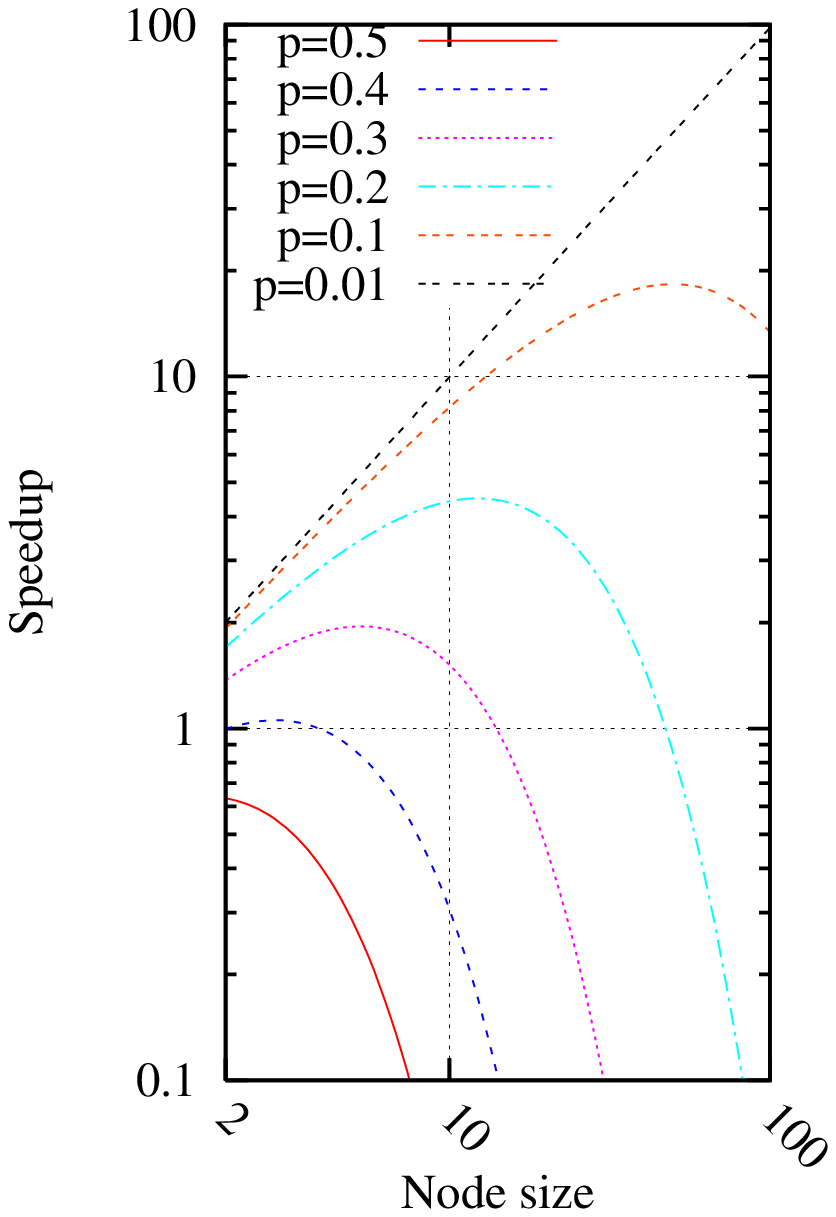}}
\subfigure[\label{simp_comm_knlogn}$c(n)=nlog_2(n)$]{\includegraphics[height=2.1in,width=2.1in]{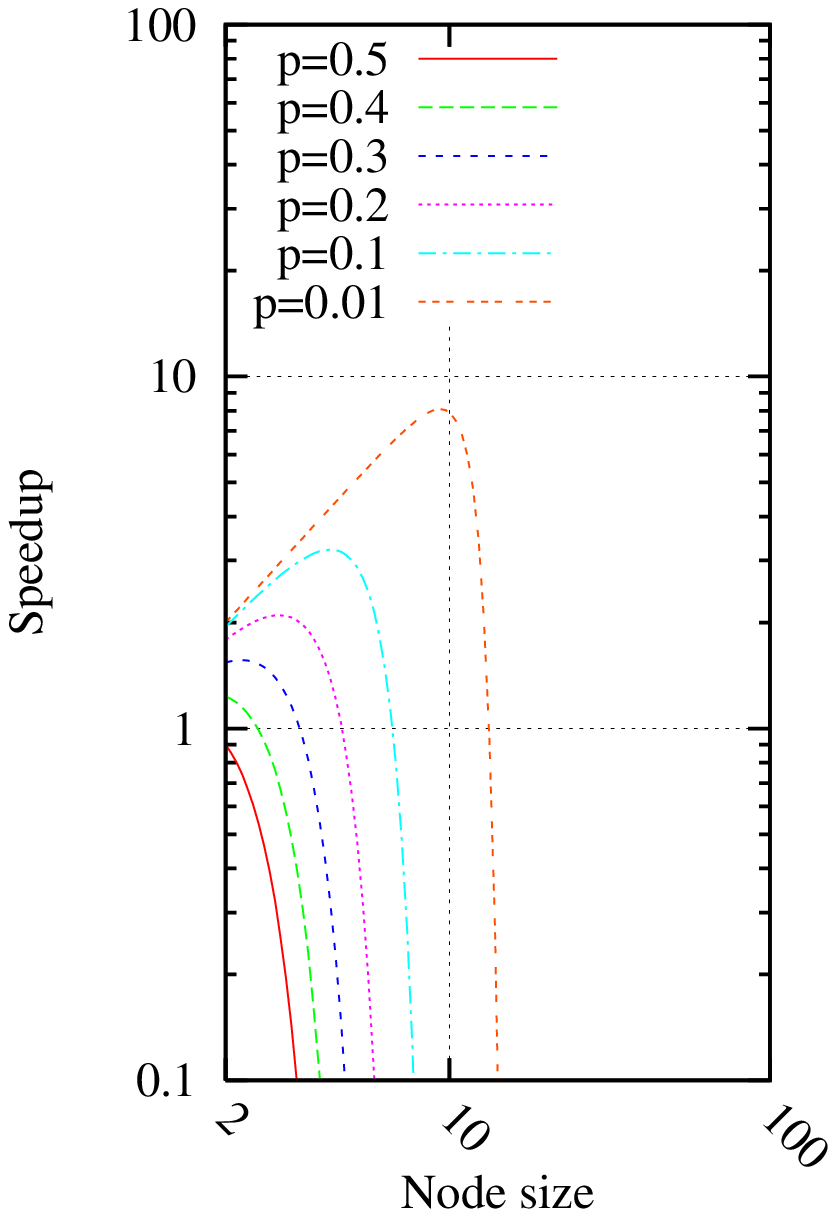}}
\subfigure[\label{simp_comm_kn2}$c(n)=n^2$]{\includegraphics[height=2.1in,width=2.1in]{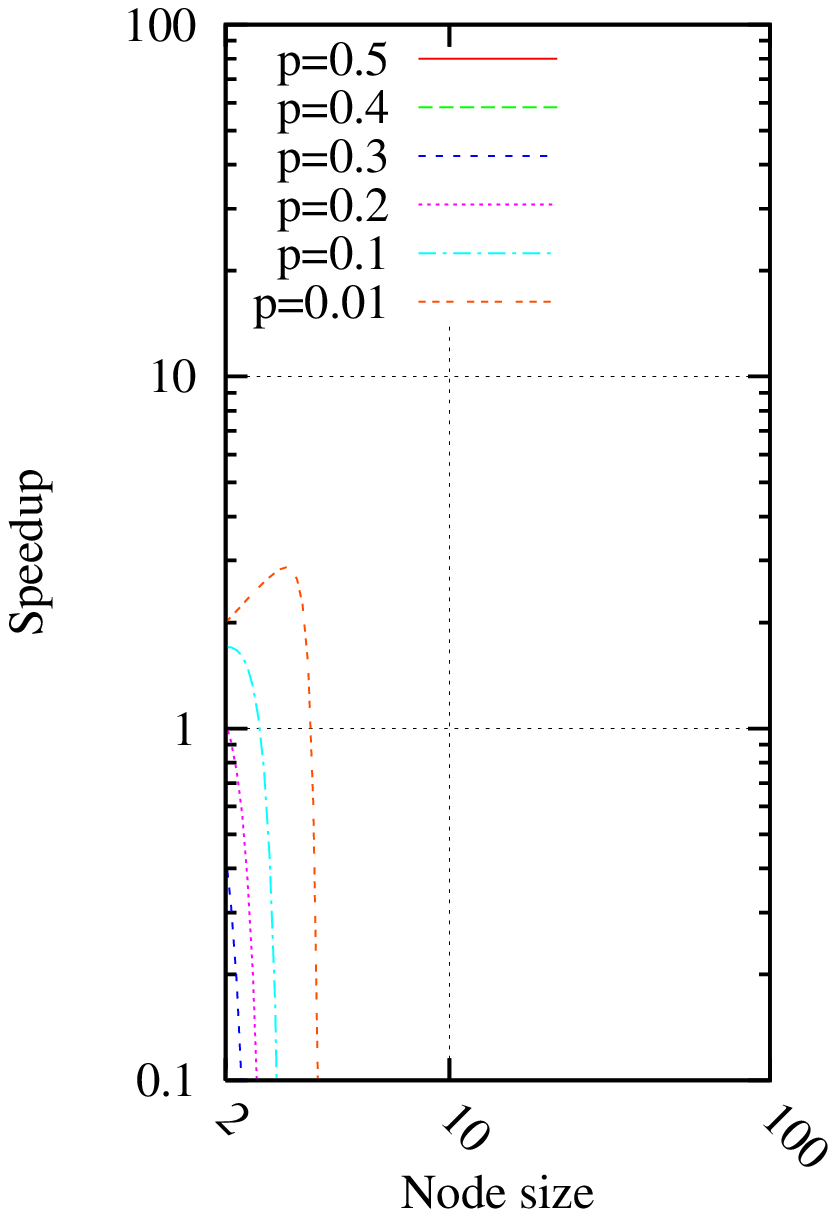}}}
\caption{\label{simp_commk}Graph depicts speedup achieved for
communication $c(n)$ at different packet loss probabilities Vs
number of processing nodes with $k=2$ for the conceptual
approach.}
\end{figure*}

If $k\geq2$ copies of the same packets are sent in each round, the
probability of success increases and is given by $p^k_s(n,p)=(1-p^{k})^{2c(n)}$.
This approach will require $kc(n)$ packets to be transmitted from
the sending nodes and it is better than when $k=1$, with $0 < p <
1$, and can be shown as below:
\begin{align}\label{prove}
p &\geq p^k, \nonumber \\
(1-p)&\leq(1-p^k), \nonumber \\
(1-p)^{2c(n)}&\leq(1-p^k)^{2c(n)}, \nonumber \\
p_s(n,p)&\leq p^k_s(n,p).
\end{align}
Transmission of $k$ copies of the same packet produces higher
probability of transmission success. Techniques used by high
performance data transfer protocols mentioned in the previous
section can be applied to reduce congestion. Their effect on the
model is not considered here. \reffig{simp_commk} shows that, when
communication is a constant e.g. $c(n)=1$, the speedup increases
linearly (note the complexity of speedup $\OO{{1}}$, e.g. a single
point to point communication in a round), when $c(n)=log_2(n)$ (e.g.
binomial tree, Bruck and recursive doubling\cite{Thakur05}
algorithm for broadcast), the speedup is monotonically increasing
with $n$, $\OO{{n^{(1-2p^k)}}}$. However, when $c(n)=log_2^2(n)$,
$c(n)=n$ (e.g. Van de Geijin\cite{Thakur05} algorithm for
broadcast and ring method for all-gather collective
communication), $c(n)=nlog_2(n)$ and $c(n)=n^2$ (e.g. naive
all-to-all algorithm) the speedup function is not monotonic and
there exists an optimal value of $n$ for a given $p$. The optimal
value gives an indication on possible number of nodes $n$ to use
when communication cost is zero. Furthermore, it also shows the
scalability of an algorithm with different type of communication
and varying packet loss probability as depicted in
\reffig{simp_commk}.
\begin{comment}
\begin{table*}[htbp]
\begin{tabular}{|c|c|c|}
  \hline
  &\multicolumn{1}{c|}{Conceptual approach }&\multicolumn{1}{c|}{L-BSP Model}\\
  \hline
  $c(n)$ &$S_E, \text{when }n>1$ & $S_E, \text{when }n>1$\\
  \hline
  $1$ & $ne^{-2p^k}$ &  $\frac{G_1n}{e^{2p^k}+G_1}$\\
  \hline
  $log(n)$ & $n^{(1-2p^k)}$ & $\frac{G_1n}{n^{2p^k}+G_1}$ \\
  \hline
  $log^2(n)$ & $ne^{-2p^klog^2(n)}$ & $\frac{G_1n}{e^{2p^klog^2(n)}+G_1}$ \\
  \hline
  $n$ & $ne^{-2p^kn}$ & $\frac{G_1n}{e^{2p^kn}+G_1}$  \\
  \hline
  $nlog(n)$& $ne^{-2p^knlog(n)}$ &  $\frac{G_1n}{e^{2p^knlog(n)}+G_1}$  \\
  \hline
  $n^2$ & $ne^{-2p^kn^2}$ & $\frac{G_1n}{e^{2p^kn^2}+G_1}$  \\
  \hline
\end{tabular}
  \centering
  \caption{\label{models}Approximate speedup for different $c(n)$ for the conceptual and L-BSP model.}
\end{table*}
\end{comment}

\begin{figure}[htbp]
\begin{center}
\psfrag{Wn}{$\frac{w}{n}$} \psfrag{T1}{$T(1)$} \psfrag{TR}{$T_R$}
\psfrag{tau}{$2\tau$} \psfrag{1}{$1$} \psfrag{2}{$2$}
\psfrag{3}{$3$} \psfrag{R}{$\hat{\rho}$}
\includegraphics[height=1.5in,width=3.3in]{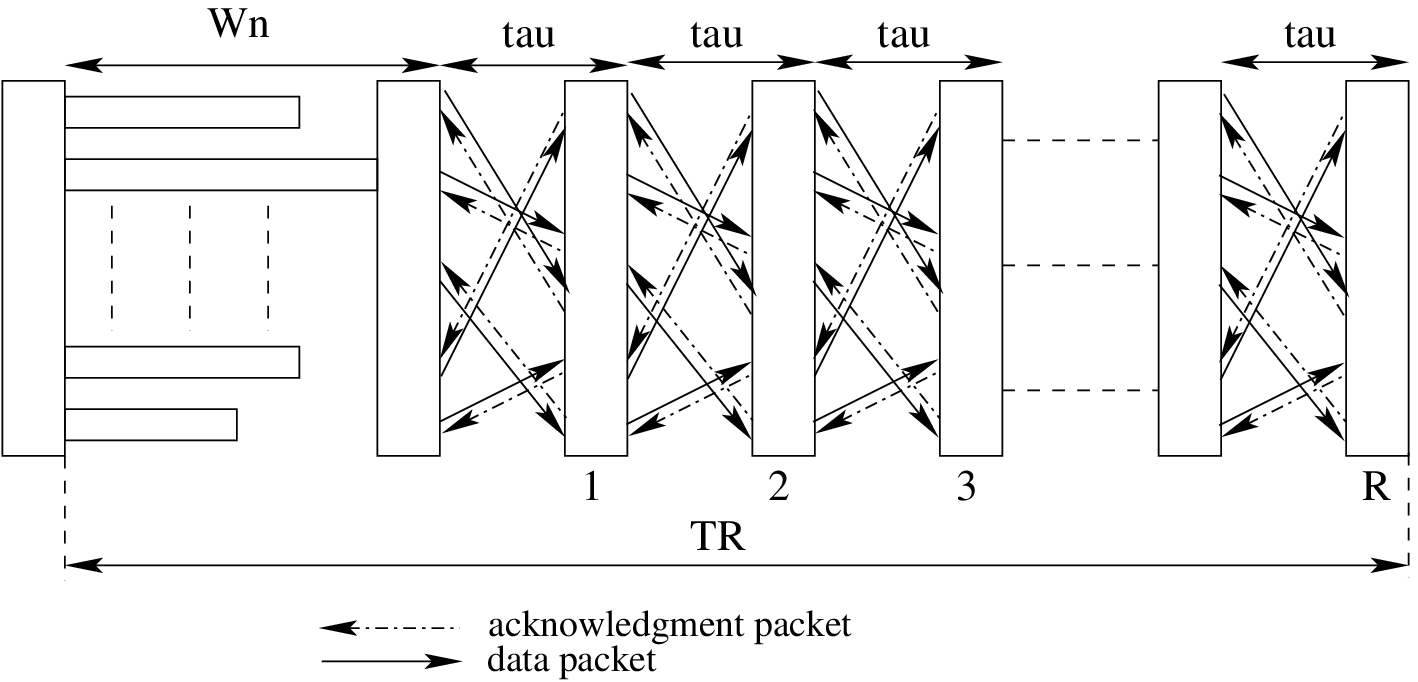}
\caption{\label{retr_BSP_UDP}Retransmission of packets using UDP
in BSP type parallel program in a superstep.}
\end{center}
\end{figure}

The conceptual approach used in this section can be simplified
further by using :
\begin{equation}\label{appr}
e^x= \lim_{b\rightarrow\infty}\bigg(1+\frac{x}{b} \bigg)^b.
\nonumber
\end{equation}
Probability of success $p_s^k(n,p)=(1-p^k)^{2c(n)}$, can be
approximated as $p_s^k(n,p)\approx e^{-2p^kc(n)}$. When
communication is $c(n)=log_2(n)$, $c(n)=log_2^{2}(n)$, $c(n)=n$,
$c(n)=nlog_2(n)$, and $c(n)=n^2$ the expected speedup can be
approximated as $S_E=ne^{-2p^kc(n)}$ when $p$ is small.
%, as shown in Table. \ref{models}.
There exists an optimal value $n$ for given $p$ when communication
$c(n)$ is not $1$ or $log_2(n)$. For $c(n)=log_2^2(n)$, $c(n)=n$, and
$c(n)=n^2$  optimal number of processors to use is $\big\lfloor
e^{\frac{\ln^22}{4p^k}}\big\rfloor$,$\big\lfloor\frac{1}{2p^k}\big\rfloor$
and $\bigg\lfloor\frac{1}{2\sqrt{p^k}}\bigg\rfloor$ respectively.
When $c(n)=nlog_2(n)$, no analytical solution exists but a numerical
solution can be found.

\section{The Lossy BSP (L-BSP) model}

In this section a model to better reflect the behavior of parallel
programs on VLSG is introduced. This model, called the Lossy BSP
(L-BSP), includes important characteristics of the internet such
as average end-to-end round-trip time and average end-to-end
bandwidth. We fix value of $2\tau$ as the timeout period for
sending data packets to and receiving acknowledgment packets from
its destination respectively, refer \reffig{retr_BSP_UDP}. $\tau$
is defined as:
\begin{equation}\label{comm_cost}
\tau =\frac{c(n)}{n}\alpha + \beta. \nonumber
\end{equation}
where $\alpha=\frac{packet\text{ }size}{bandwidth}$ and $\beta$,
the delay is the round trip time (includes cost for sending data
and receiving acknowledgment packet). $T(1)=wr$ is the time taken
for computation in one processor, with $r$ as the number of times
computation is repeated (known as supersteps in BSP model).
Whereas, $T_R=\frac{w}{n}+2\hat{\rho}\tau$ refers to the expected
time that $n$ processors will take to complete a single round as
shown in \reffig{retr_BSP_UDP}. With zero packet loss (i.e.
$\hat{\rho}=1$), time $T(n,\tau)=\frac{T(1)}{n}+2r\tau$ for
computation and communication on $n$ nodes can be achieved.
Ideally, we expect speedup $S_I=\frac{T(1)}{T(n,\tau)}=n$,
assuming zero communication cost (i.e. $\tau=0$). However, with an
expected average number of transmission
$\hat{\rho}=\frac{1}{p_s(n,p)}$ (all packets are re-transmitted if
a packet is lost), the expected time taken for computation and
communication is given by
$\hat{T}(n,p,\tau)=\frac{T(1)}{n}+\frac{2r\tau}{p_s(n,p)}$. Hence,
the expected speedup achievable is:
\begin{equation}\label{speedup_amdahl}
S_E=\frac{wr}{\frac{T_1}{n}+\frac{2r\tau}{p_s(n,p)}}=\frac{w}{\frac{w}{n}+\frac{2\tau}{p_s(n,p)}}=\frac{nGp_s(n,p)}{1+Gp_s(n,p)},\nonumber
\end{equation}
where granularity (ratio of computation and communication costs)
is $G=\frac{w}{2n\tau}$.

If only lost packets are re-transmitted, data packets that have
been successfully sent will not be re-transmitted again.
Therefore, the sequence of packet transmission is given by $c(n)$,
$pc(n)$, $p^2c(n)$, $p^3c(n)$,$\cdots$. That is, in the first
transmission $c(n)$ packets are sent, in the second transmission
only packets that are lost (i.e. $pc(n)$) are re-transmitted and
so on. The notion used in \reffeq{N_of_transm} can also be applied
in this model. If $p_s=(1-p)^2$ is the probability of successful
delivery of a single packet and $c(n)$ is the number of packets
transmitted then the probability that a communication terminates
in the $i$-th re-transmission is $(1-p_s)^{(i-1)}p_s$ for a single
packet. Thus, $\sum_{j=1}^{i}(1-p_s)^{(j-1)}p_s$ is the
probability that a communication terminates by $i$th
re-transmission and
$\big[\sum_{j=1}^{i}(1-p_s)^{(j-1)}p_s\big]^{c(n)}$ is the
probability that all communications terminate by $i$-th
re-transmission. It follows that, the average number of
re-transmission can be obtained from:
\begin{align}
\label{transmit_only_lost_p} &\hat{\rho}(p_s,c(n))
=\sum_{i=1}^{\infty}i\bigg(\bigg[\sum_{j=1}^{i}(1-p_s)^{(j-1)}p_s
\bigg]^{c(n)}- \nonumber \\
&\bigg[\sum_{j=1}^{i-1}(1-p_s)^{(j-1)}p_s\bigg]^{c(n)}\bigg)
\nonumber \\
&=\sum_{i=1}^{\infty}i\bigg(\bigg[1-(1-p_s)^i\bigg]^{c(n)} -
\bigg[1-(1-p_s)^{i-1}\bigg]^{c(n)}\bigg)\nonumber \\
\end{align}

\begin{figure*}[htbp]
\centerline{\subfigure[\label{SE_n_W14400_logn}$c(n)=1$]{\includegraphics[height=2.1in,width=2.1in]{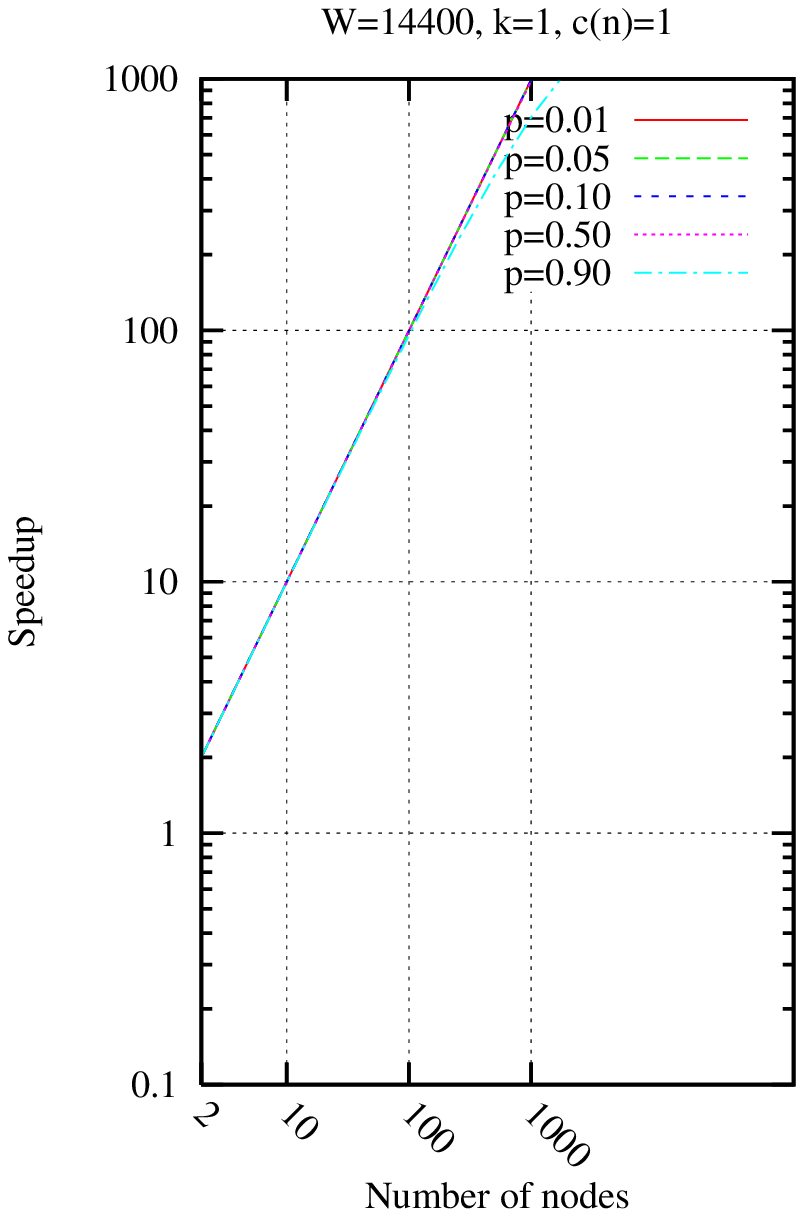}}
\subfigure[\label{SE_n_W14400_logn}$c(n)=log_2(n)$]{\includegraphics[height=2.1in,width=2.1in]{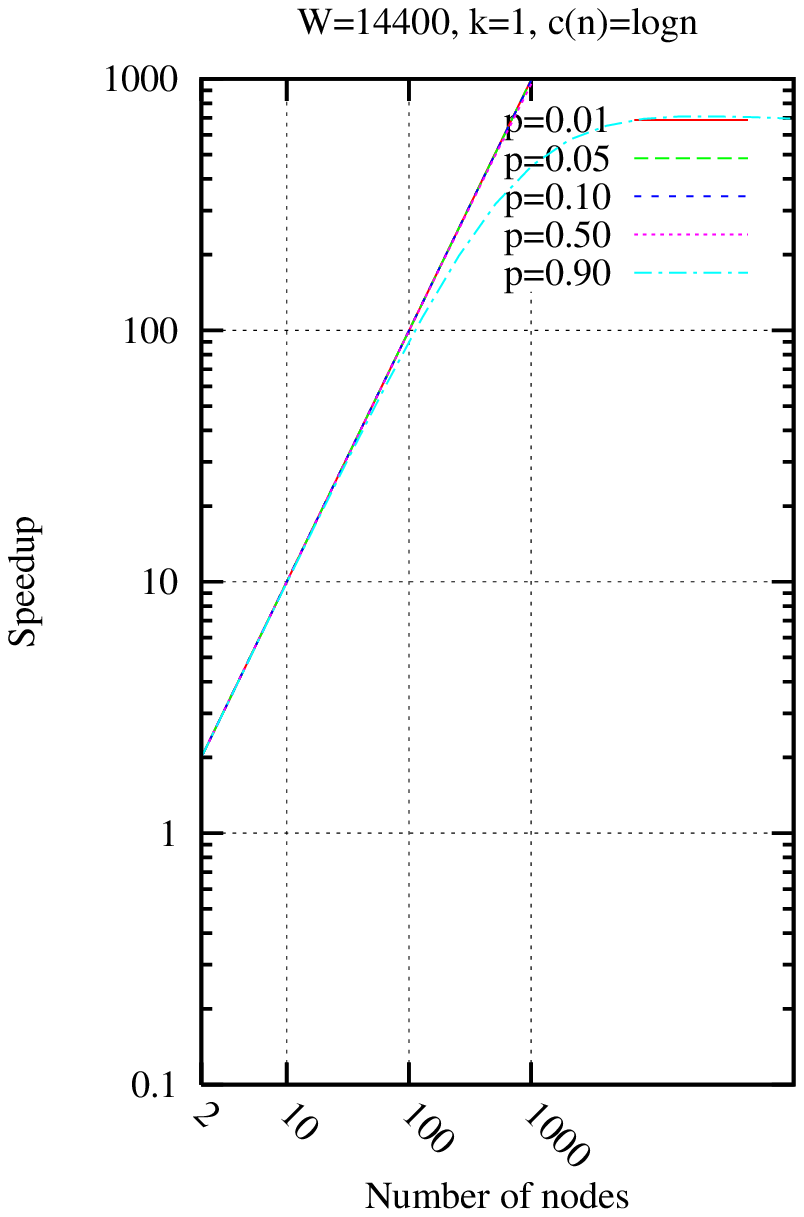}}
\subfigure[\label{SE_n_W14400_log2n}$c(n)=log_2^2(n)$]{\includegraphics[height=2.1in,width=2.1in]{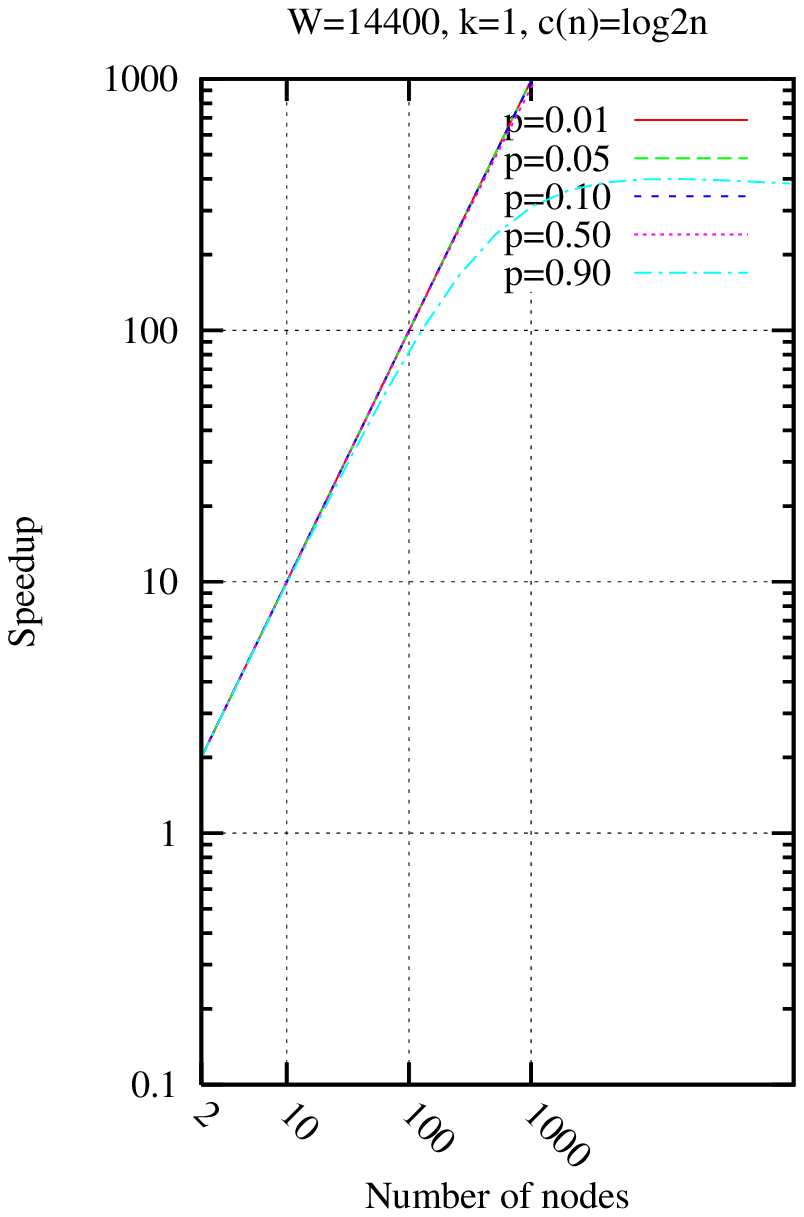}}}
\centerline{\subfigure[\label{SE_n_W14400_n}$c(n)=n$]{\includegraphics[height=2.1in,width=2.1in]{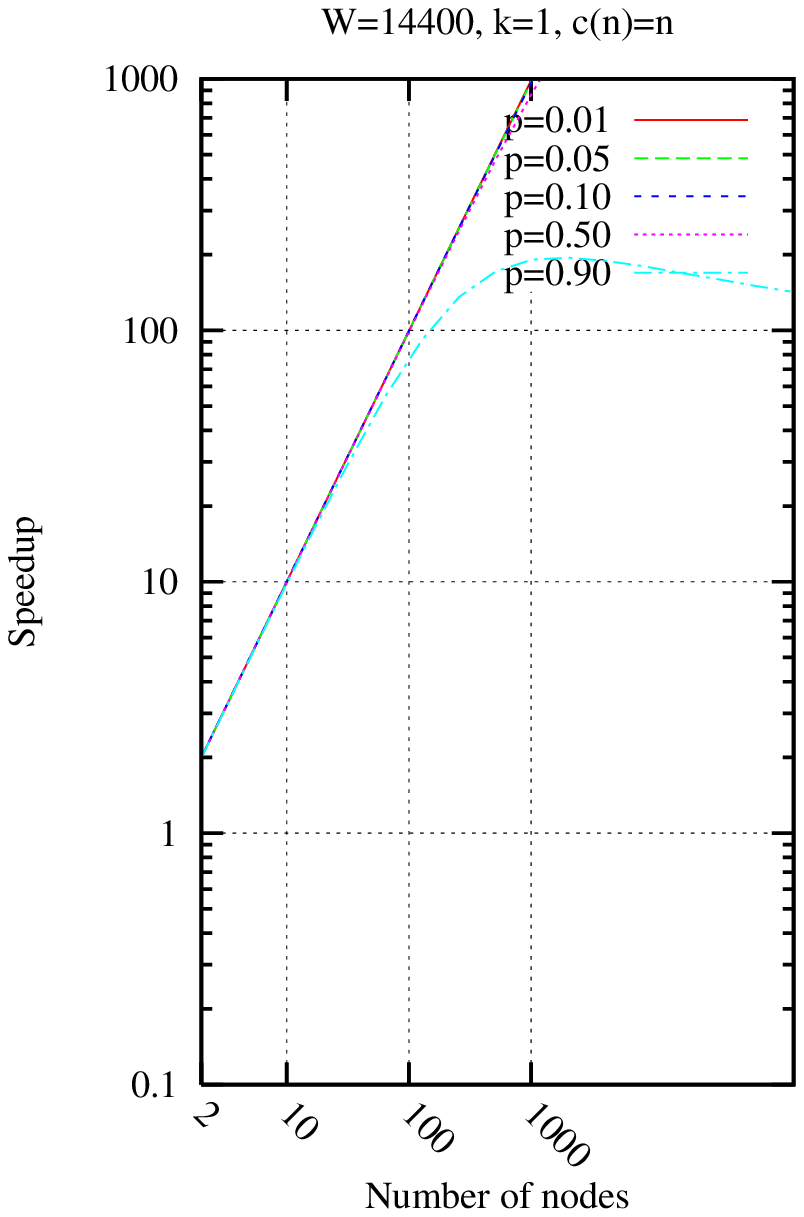}}
\subfigure[\label{SE_n_W14400_nlogn}$c(n)=nlog_2(n)$]{\includegraphics[height=2.1in,width=2.1in]{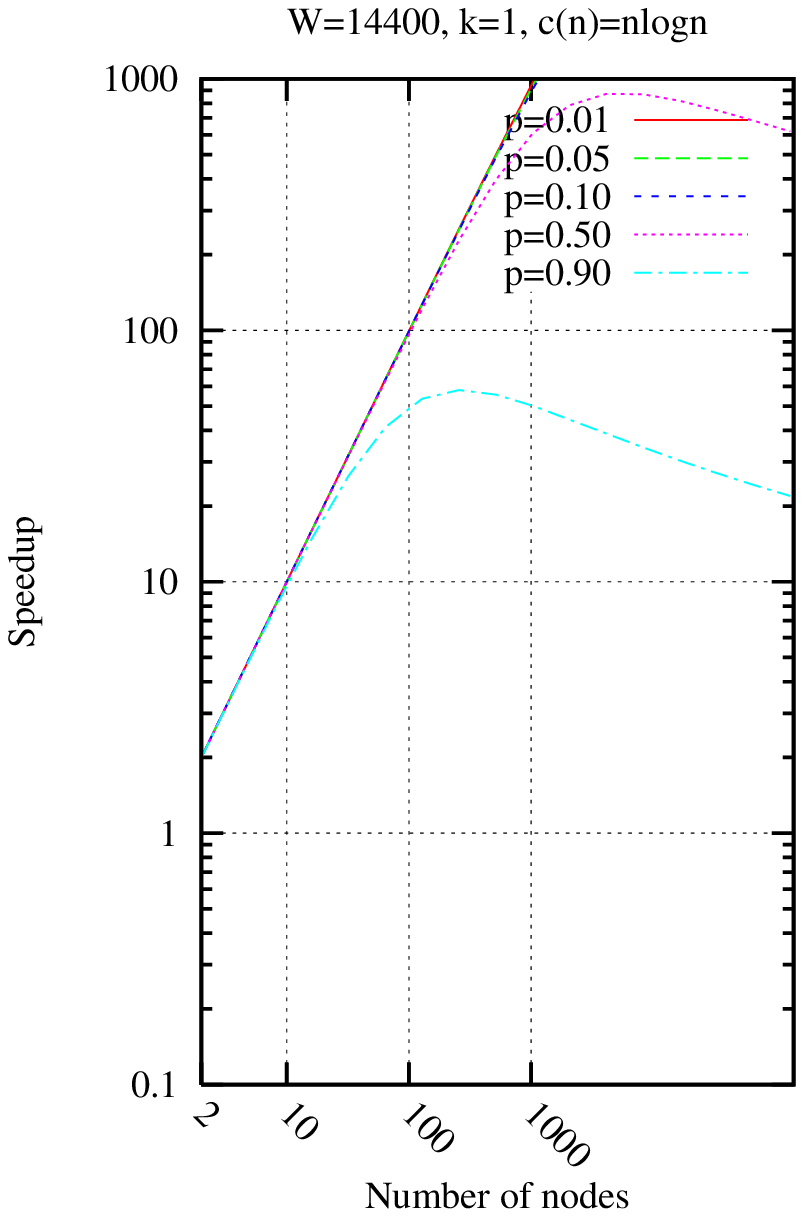}}
\subfigure[\label{SE_n_W14400_n2}$c(n)=n^2$]{\includegraphics[height=2.1in,width=2.1in]{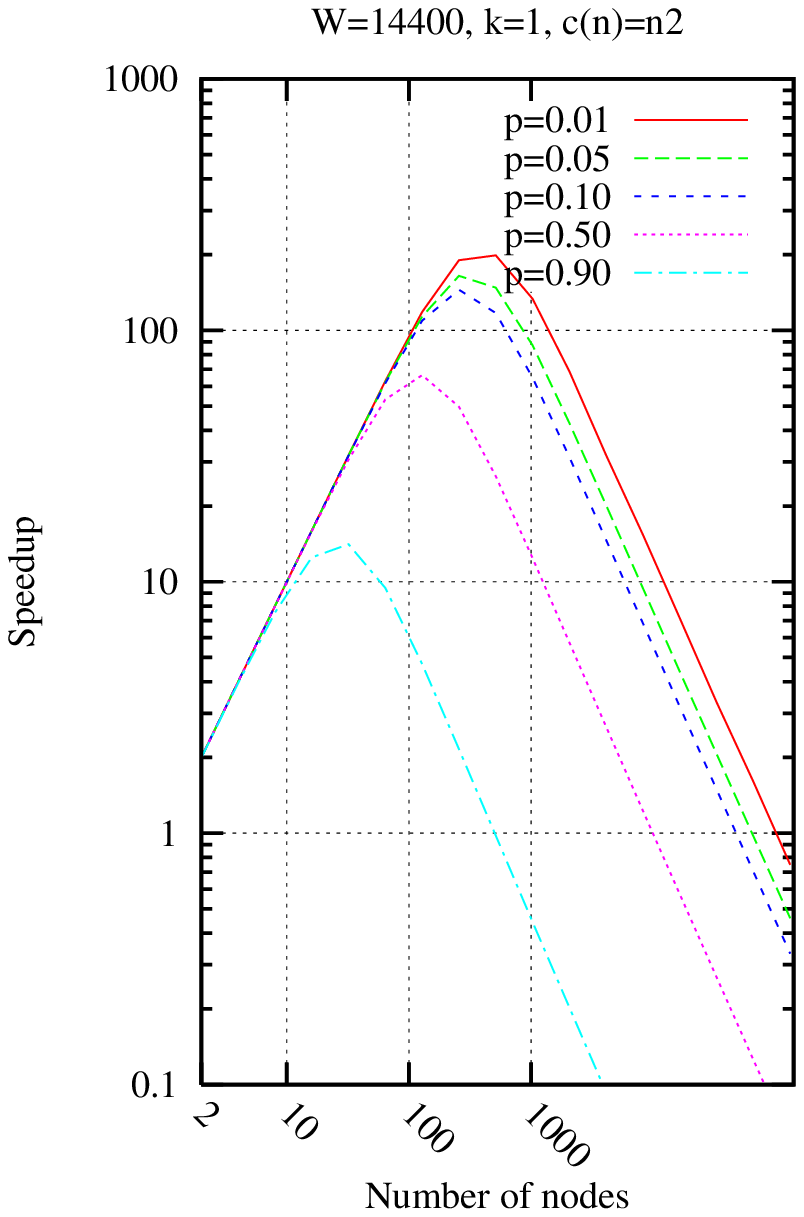}}}
\caption{\label{SE_n_W14400}Graph depicts speedup achieved for
different number of nodes $n$, work $W=4$ hours, communication
$c(n)$ at different packet loss probability with $k=1$ for the
L-BSP model.}
\end{figure*}

The value of $\hat{\rho}(p_s,c(n))$ depends on packet loss $p$ and
communication $c(n)$. We use numerical approach to obtain values
of $\hat{\rho}(p_s,c(n))$ for different number of nodes. Thus the
expected speedup obtained using the L-BSP model can be simplified
as:
\begin{equation}
\label{speedup_transmit_only_lost_p}
S_E=\frac{Gn}{G+\hat{\rho}{(p_s,c(n))}}
\end{equation}
with granularity, $G=\frac{w}{2n\tau}$. Clearly, speedup
approaches linearity when $G\gg\hat{\rho}(p_s,c(n))$.
\reffig{SE_n_W14400} shows the effect of granularity for different
communication $c(n)$. For higher communication complexity such as
$c(n)=nlog_2(n)$ and $c(n)=n^2$, \reffig{SE_n_W14400_nlogn} and
\reffig{SE_n_W14400_n2} respectively, speedup deteriorates at a
faster rate.

\begin{figure*}[htbp]
\centerline{\subfigure[\label{SE_p_W36000_1}$c(n)=1$]{\includegraphics[height=2.1in,width=2.1in]{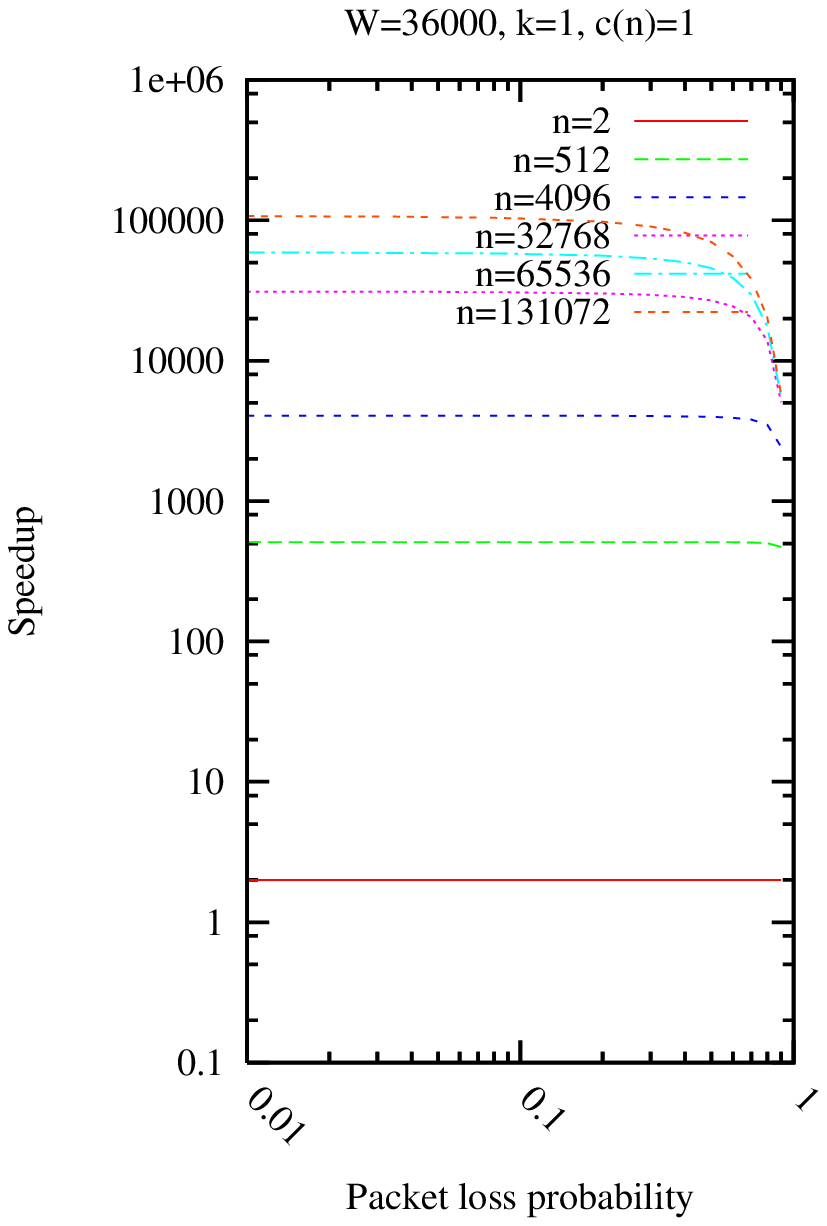}}
\subfigure[\label{SE_p_W36000_logn}$c(n)=log_2(n)$]{\includegraphics[height=2.1in,width=2.1in]{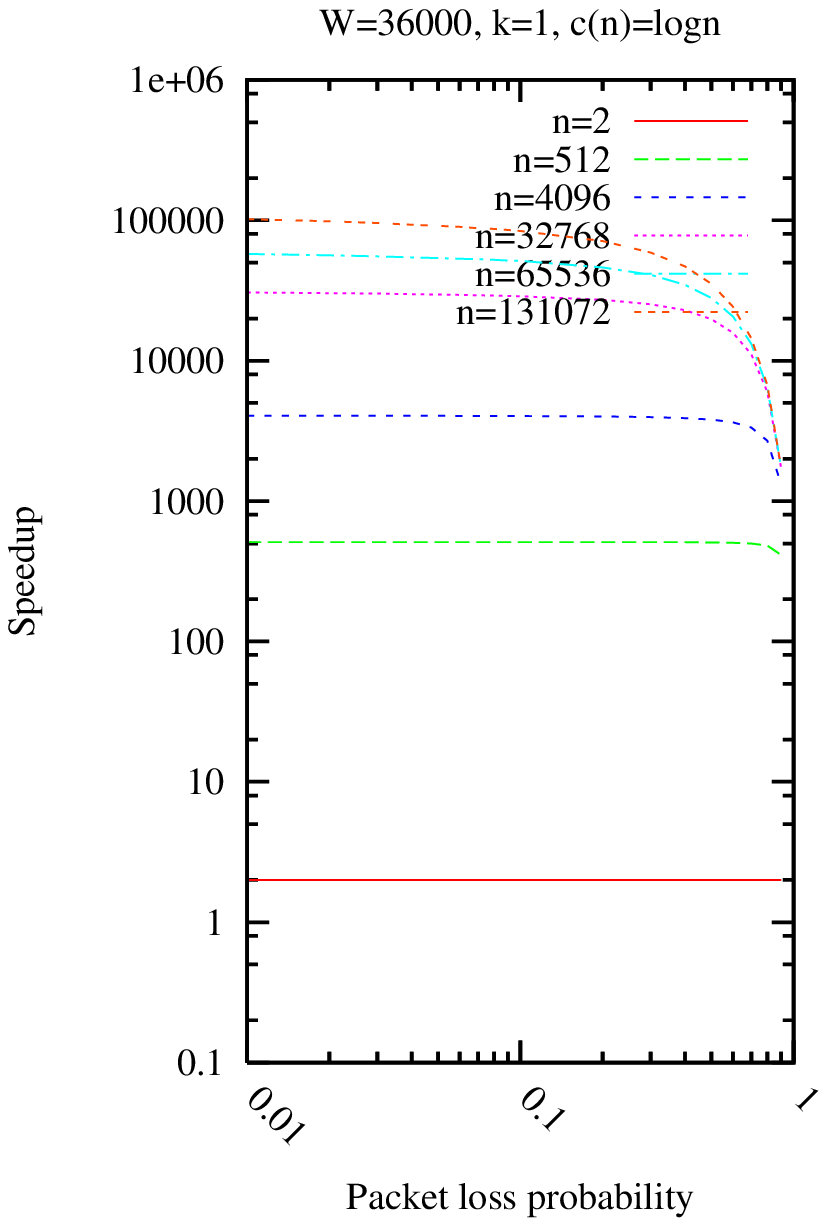}}
\subfigure[\label{SE_p_W36000_log2n}$c(n)=log_2^2(n)$]{\includegraphics[height=2.1in,width=2.1in]{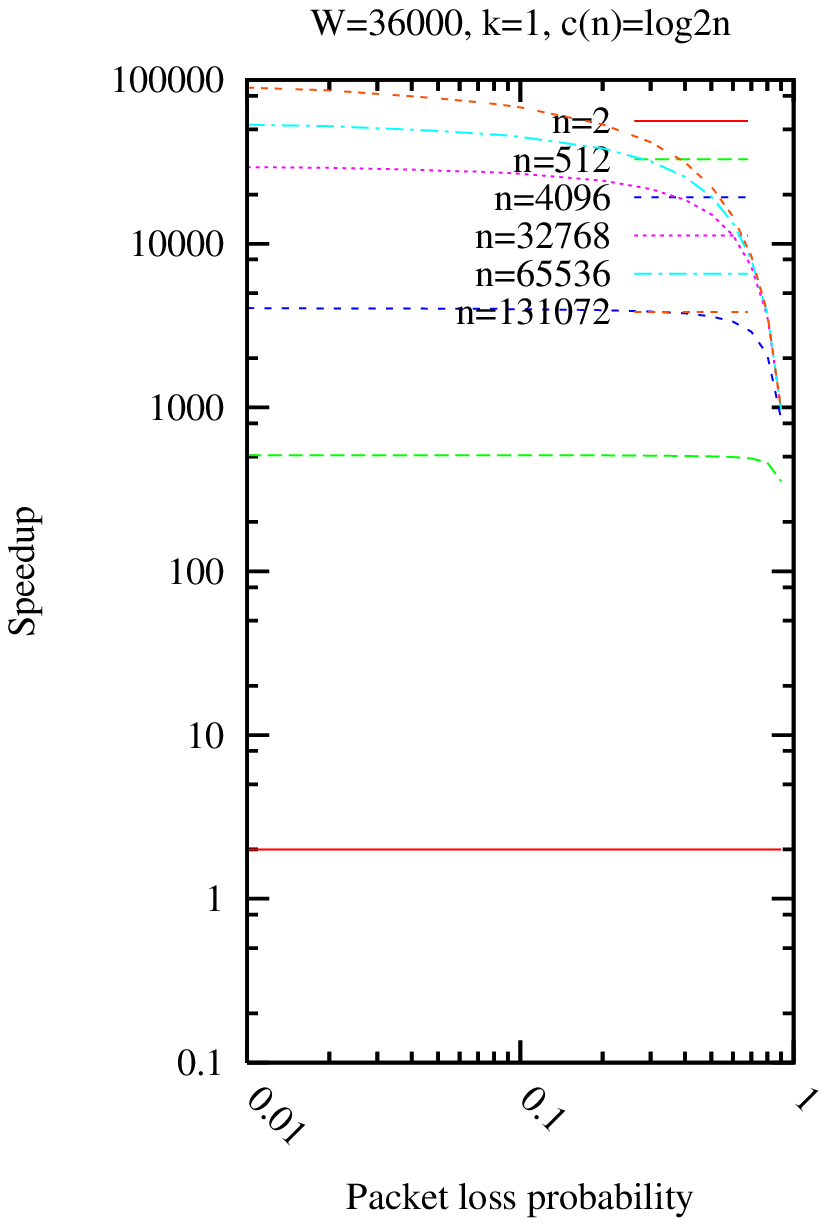}}}
\centerline{\subfigure[\label{SE_p_W36000_n}$c(n)=n$]{\includegraphics[height=2.1in,width=2.1in]{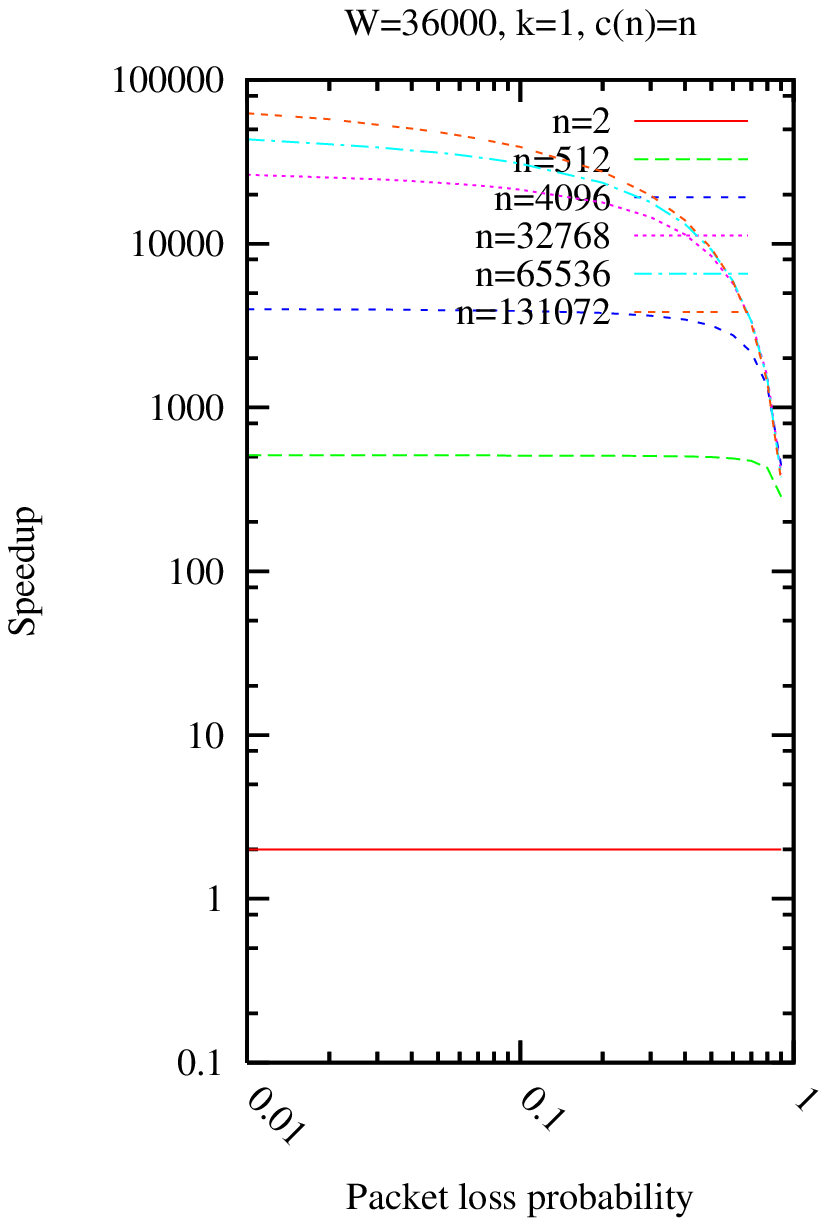}}
\subfigure[\label{SE_p_W36000_nlogn}$c(n)=nlog_2(n)$]{\includegraphics[height=2.1in,width=2.1in]{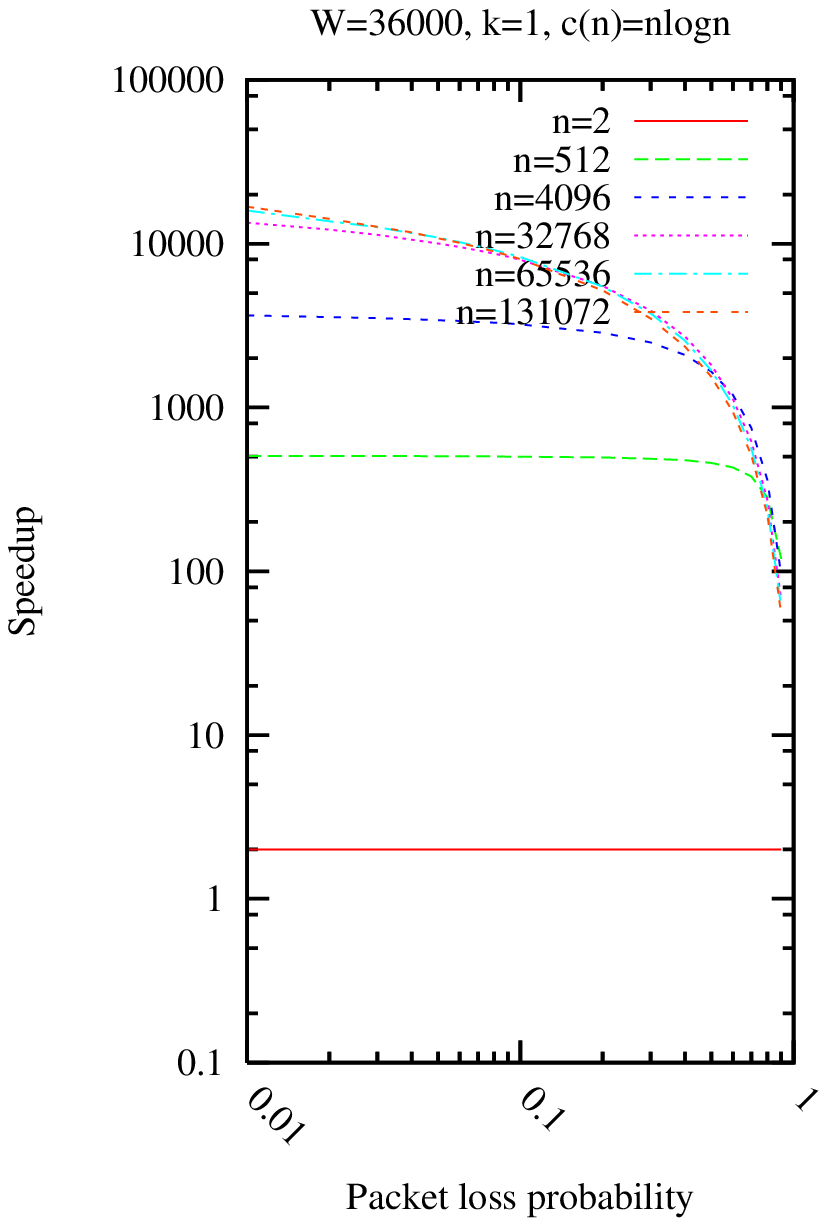}}
\subfigure[\label{SE_p_W36000_n2}$c(n)=n^2$]{\includegraphics[height=2.1in,width=2.1in]{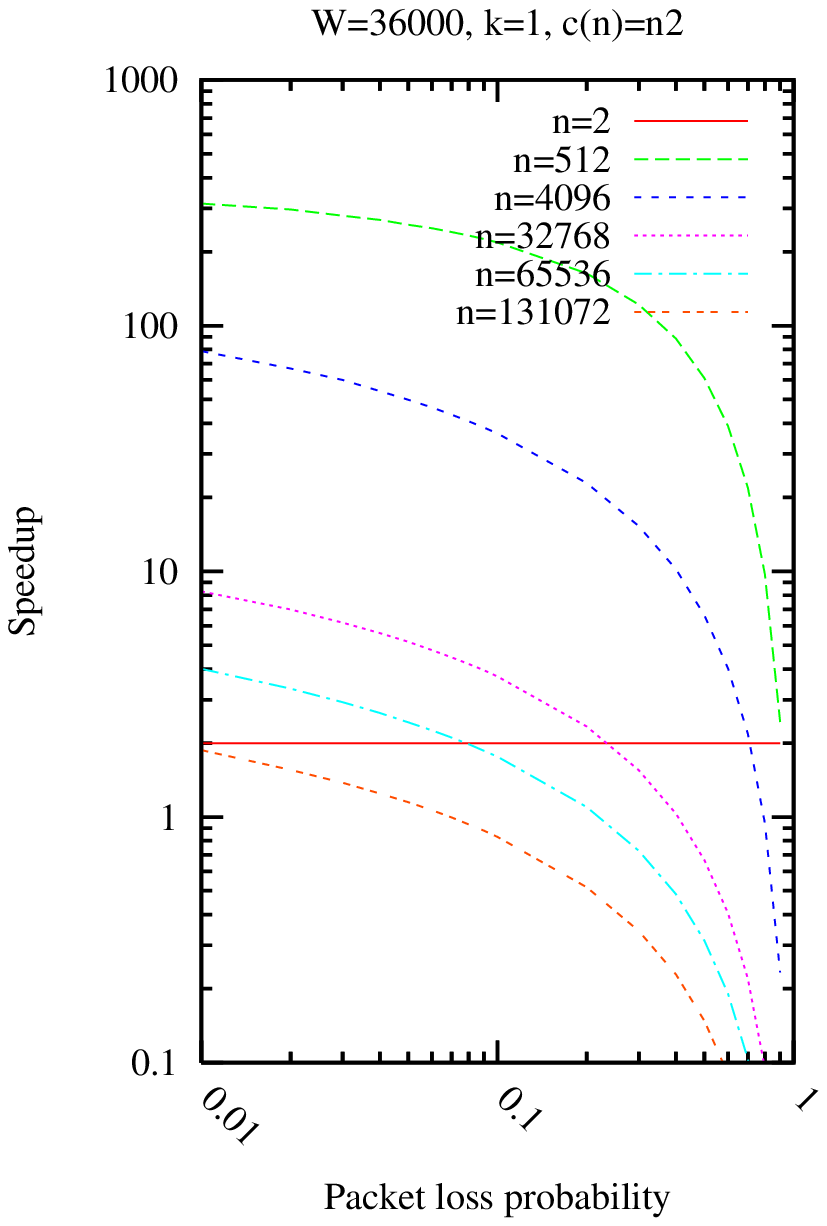}}}
\caption{\label{SE_p_W36000}Graph depicts speedup for different
packet loss probabilities given $n$ nodes, work $W=10$ hours, and
communication $c(n)$ for the L-BSP model.}
\end{figure*}

\reffig{SE_p_W36000} depicts the limits of speedup for different
probability of packet losses when different number of nodes are
used. It shows that when packet loss is lower, higher speedup can
be achieved. Number of optimal nodes to use depend on different
operating conditions such as computational and communication
complexity. When packet loss is higher, speedup deteriorates at a
faster rate. On the other hand, it demonstrates the important
effect of granularity on speedup. Linear speedup is possible with
higher granularity and correct number of nodes, this is true even
for higher degree of communication complexity and packet loss.
(e.g. $n=2$)

%and can be simplified into the following equation:
%\begin{equation}\label{t_losspacket}
%\hat{\rho}=\sum_{i=1}^{\infty}i\big[p_s(n,p)\big]^{p^{i-1}}\prod_{j=1}^{i-1}(1-\big[p_s(n,p)\big]^{p^{j-1}}).
%\end{equation}
%The value of $\rho$ is a function of $p$ and $c(n)$.

\section{Optimal packet copies}
The conceptual approach revealed that speedup can be increased by
transmitting multiple copies of the same packet. It is easy to
verify that as $k\rightarrow\infty$, we have $p_s\rightarrow 1$,
however, it is unrealistic to send that many packet copies. Thus,
finding optimal value for $k$ is necessary for a given value of
$p$, $n$, $w$ and $c(n)$.

We consider this scenario in the L-BSP model. If $k$ copies of
same packets are sent, \reffeq{speedup_transmit_only_lost_p}, can
be re-written as:
\begin{equation}\label{speedup_amdahl_kcopies}
S_E=\frac{nG_1}{G_1+\hat{\rho}^k(p_s^k,c(n))},
\end{equation}
with $p_s^k=(1-p^k)^2$, $\hat{\rho}^k(p_s^k,c(n))$ the average
number of transmissions when $k$ packet copies are used, and
$G_1=\frac{w}{2n\tau_k}$ where $\tau_k$ is defined as
$\tau_k=k\frac{c(n)}{n}\alpha + \beta$ and $2\tau_k$ represents
the timeout value for sending $kc(n)$ packets.

Equation \reffeq{speedup_amdahl_kcopies} can be simplified to:
\begin{equation}\label{k_copies_amdahlspdup}
S_E=\frac{n}{1+
\frac{2\hat{\rho}^kn}{w}\bigg(k\frac{c(n)}{n}\alpha+\beta \bigg)}=
\frac{n}{1+\frac{2k\hat{\rho}^kc(n)\alpha}{w}
+\frac{2n\beta\hat{\rho}^k}{w}}.
\end{equation}
%Table. \ref{models} shows, the approximate speedup achievable for different $c(n)$.
Assuming communication $c(n)=n^2$, it is clear that the second
term in the denominator,$\frac{2k\hat{\rho}^kn^2\alpha}{w}$, grows
quadratically as $n$ increases in \reffeq{k_copies_amdahlspdup}.
Using the numerically solved value for $\hat{\rho}^k$ we find the
optimal value of $k$, by minimizing the product of
$k\hat{\rho}^k$. \reftable{LBSPmodel} shows the dominating term
that effects the speedup as $n\rightarrow\infty$ for different
communication $c(n)$. For lower communication complexity, such as
$c(n)=1,log_2^2(n)$ and $log(n)$, the dominating term as $n$
increases is $\frac{2n\beta\hat\rho^k}{w}$.

\begin{table*}[htbp]
  \centering
  \caption{\label{LBSPmodel}Dominating term for speedup using different type of communications.}
\begin{tabular}{|c|c|c|}
  \hline
  Case &\multicolumn{1}{c|}{Communication $c(n)$ }&\multicolumn{1}{c|}{Dominating term as $n\rightarrow\infty$}\\
  \hline
  I & $n^2$ & $\frac{2k\hat{\rho}^kc(n)\alpha}{w}$\\
  \hline
  II & $nlog_2(n)$ & $\frac{2k\hat{\rho}^kc(n)\alpha}{w}$\\
  \hline
  III & $n$ & $\frac{2k\hat{\rho}^kc(n)\alpha}{w}+\frac{2n\beta\hat{\rho}^k}{w}$\\
  \hline
  IV & $log_2^2(n)$ & $\frac{2n\beta\hat{\rho}^k}{w}$ \\
  \hline
  V & $log_2(n) $& $\frac{2n\beta\hat{\rho}^k}{w}$\\
  \hline
   VI & $1$ & $\frac{2n\beta\hat{\rho}^k}{w}$\\
  \hline
\end{tabular}

\end{table*}

As the time to transmit the packet, $\alpha$, approaches zero
(i.e. transmission cost approaches zero)
\reffeq{k_copies_amdahlspdup} can be reduced to:
\begin{equation}\label{reduced_k_speedup}
\lim_{\substack{k\rightarrow\infty \\ \alpha \rightarrow 0}}
S_E=\frac{n}{\frac{2n\beta}{w}+1}.\nonumber
\end{equation}
Here, $\hat{\rho}^k \rightarrow 1$, as number of packet copies,
$k$, transmitted increases. It indicates that work performed on
each node should be large enough compared to the average delay
between nodes to achieve good speedup.

\begin{comment}
\reffig{n_k_W1}, \reffig{n_k_W14400}, and \reffig{n_k_W36000}
shows the number of packet copies that could be transmitted to
achieve best possible speedup for different size of nodes, when
work, $w$, is $1$ second, $4$ hours and $10$ hours respectively
for different packet loss probability. Generally, the graphs
indicate that as number of nodes increases, the number of packet
copies required grows in tandem. However, graphs with higher
communication complexity and higher packet loss probability shows
some discrepancies. This is because of inaccuracy in computation
of $\hat{\rho}^k$ due to truncation error.
\end{comment}

\begin{figure*}[htbp]
\centerline{\subfigure[\label{SE_k_W36000_1}$c(n)=1$]{\includegraphics[height=2.1in,width=2.1in]{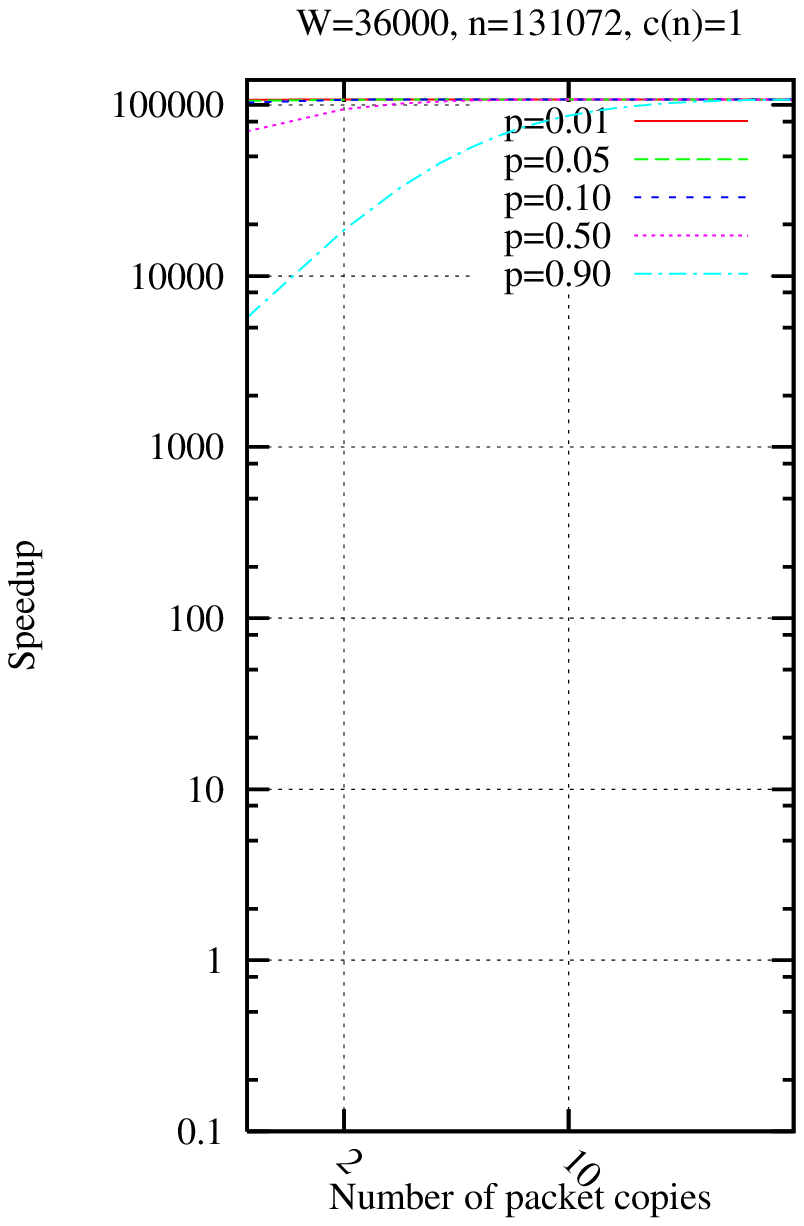}}
\subfigure[\label{SE_k_W36000_logn}$c(n)=log_2(n)$]{\includegraphics[height=2.1in,width=2.1in]{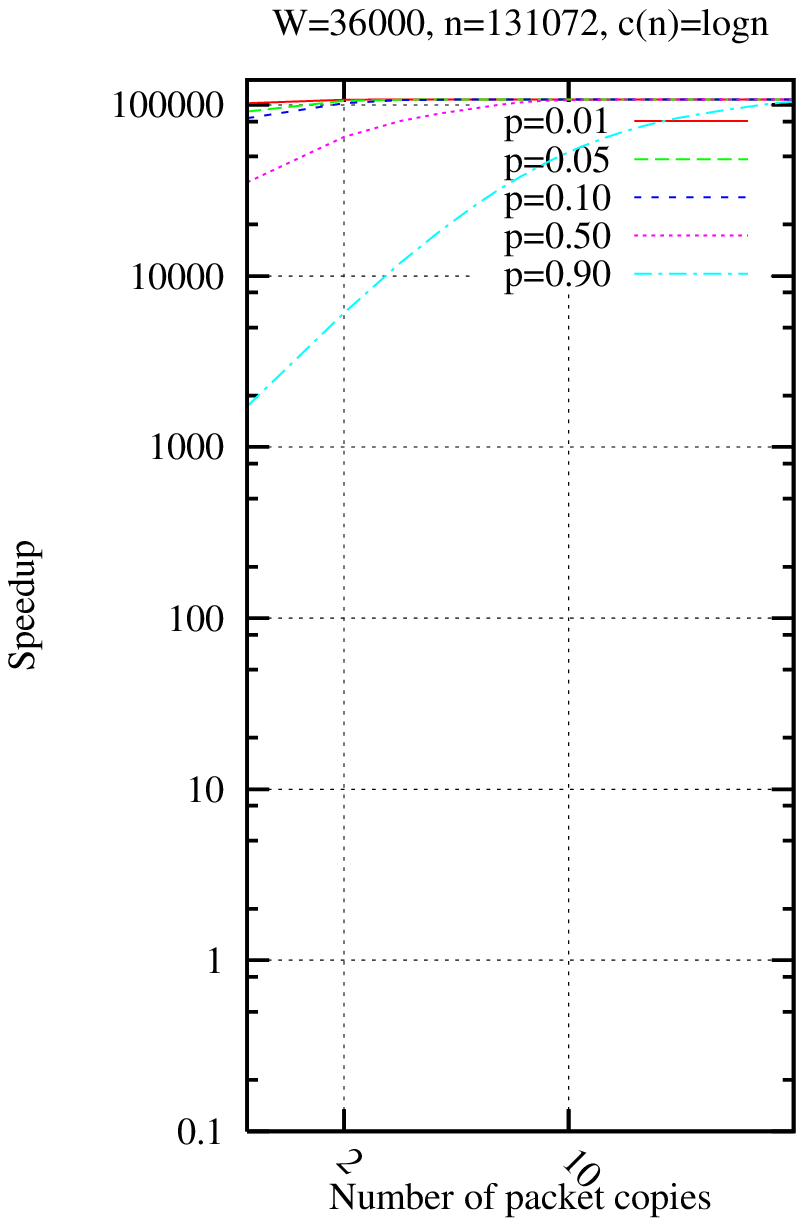}}
\subfigure[\label{SE_k_W36000_log2n}$c(n)=log_2^2(n)$]{\includegraphics[height=2.1in,width=2.1in]{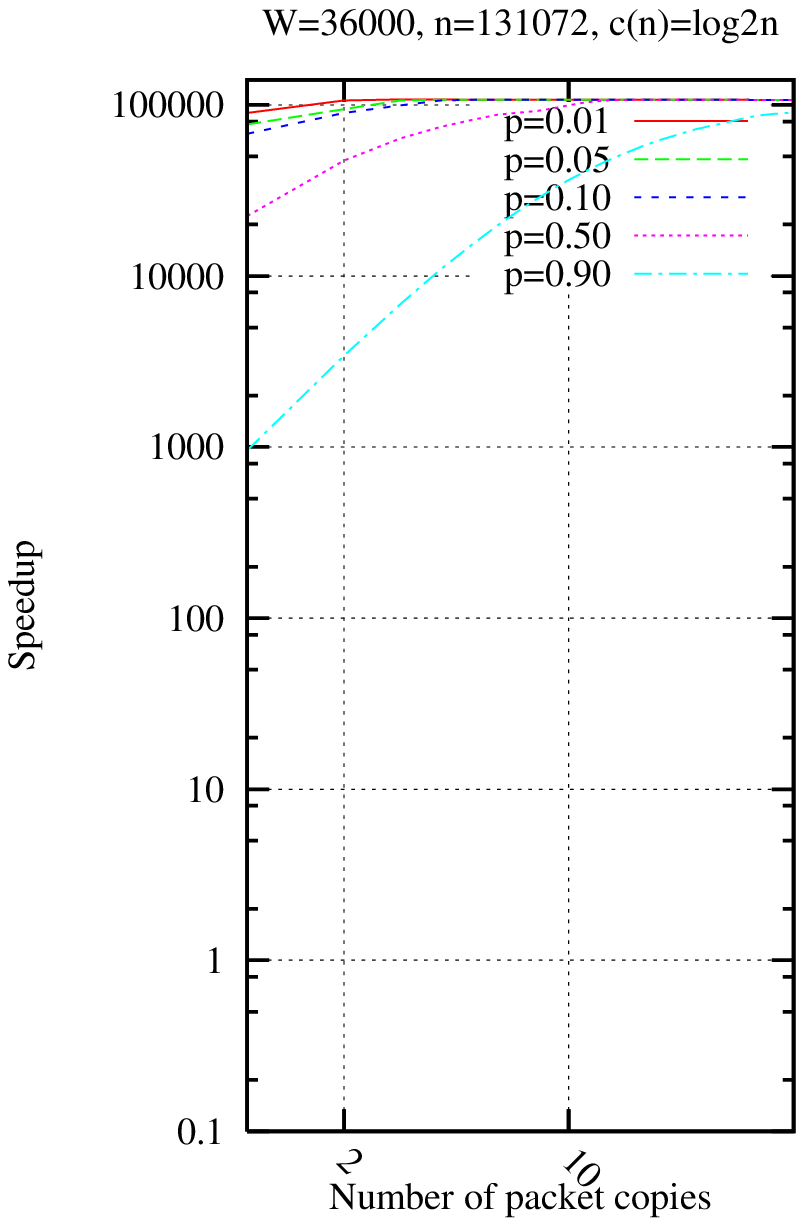}}}
\centerline{\subfigure[\label{SE_k_W36000_n}$c(n)=n$]{\includegraphics[height=2.1in,width=2.1in]{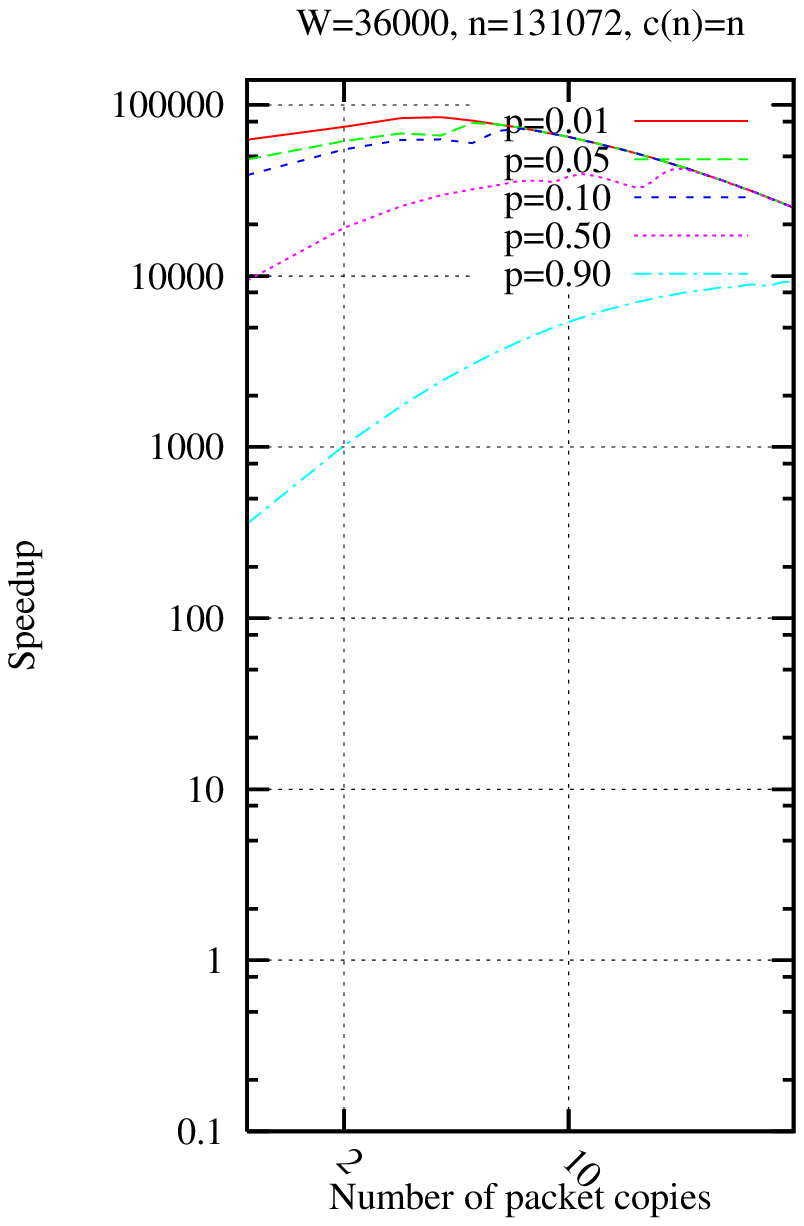}}
\subfigure[\label{SE_k_W36000_nlogn}$c(n)=nlog_2(n)$]{\includegraphics[height=2.1in,width=2.1in]{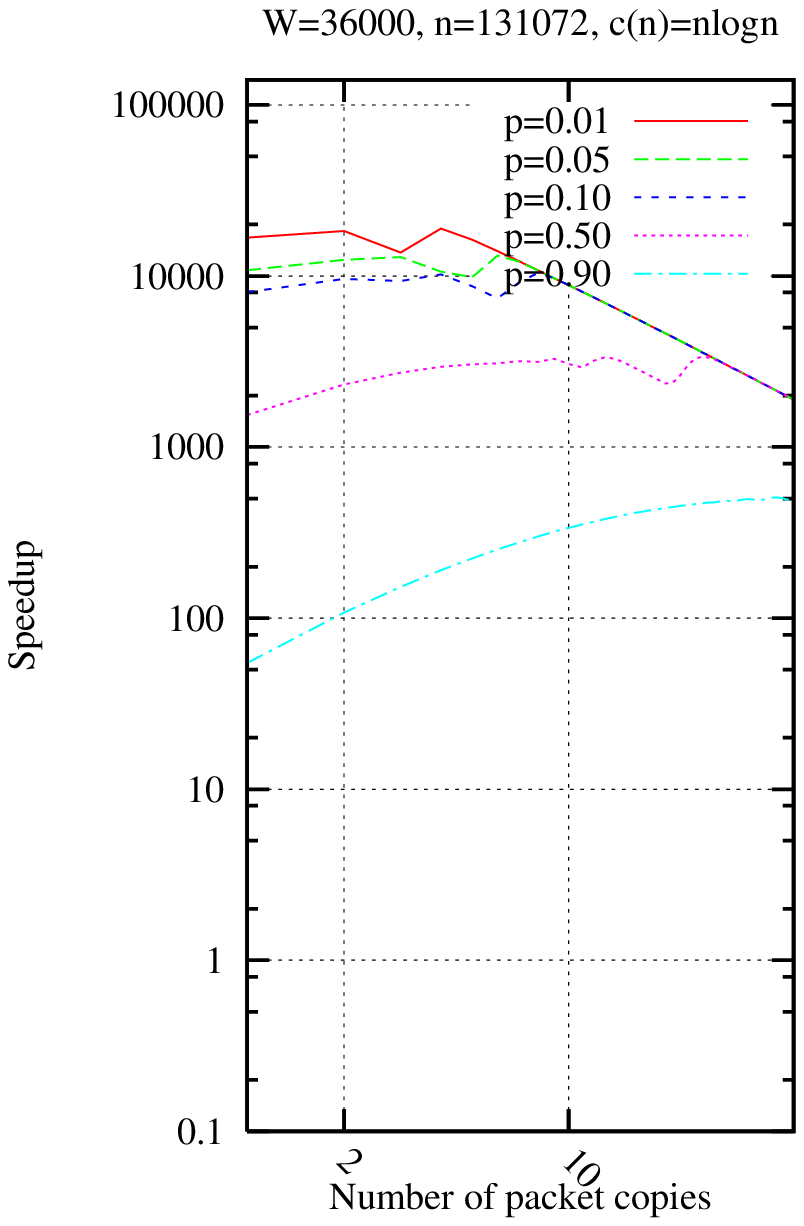}}
\subfigure[\label{SE_k_W36000_n2}$c(n)=n^2$]{\includegraphics[height=2.1in,width=2.1in]{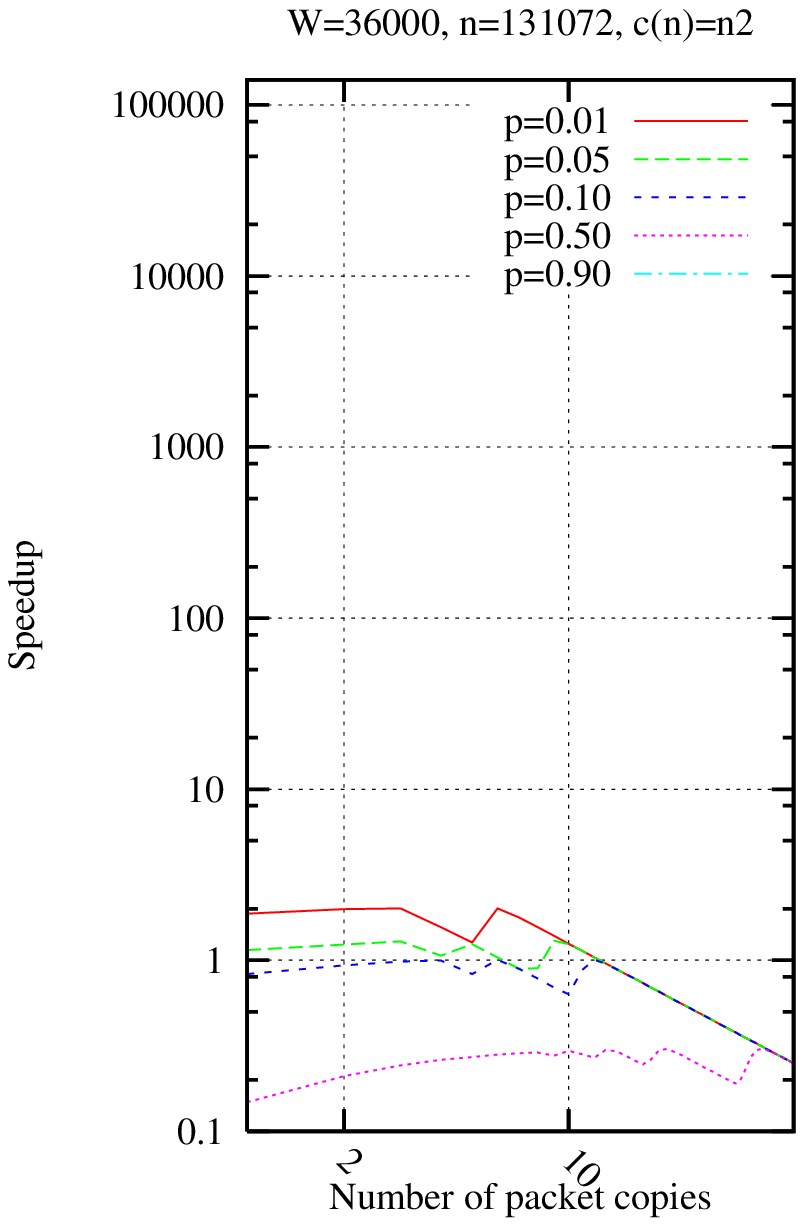}}}
\caption{\label{SE_k_W36000}Graph depicts speedup for different
packet copies given $n$ nodes, work $W=10$ hours, communication
$c(n)$ at different packet loss probability for the L-BSP model.}
\end{figure*}

\reffig{SE_k_W36000} shows how speedup is effected by the number
of packet copies transmitted for work of $w=10$ hours. It is clear
from the figure that for communication $c(n)=n$, $c(n)=nlog_2(n)$
and $c(n)=n^2$ speedup deteriorates as the number of packet copies
used increases. This observation concurs with the expected
decrease in speedup, because of higher overhead caused by more
packets and higher communication complexity.

\begin{figure*}[htbp]
\centerline{\subfigure[\label{SE_W_n2_1}$c(n)=1$]{\includegraphics[height=2.1in,width=2.1in]{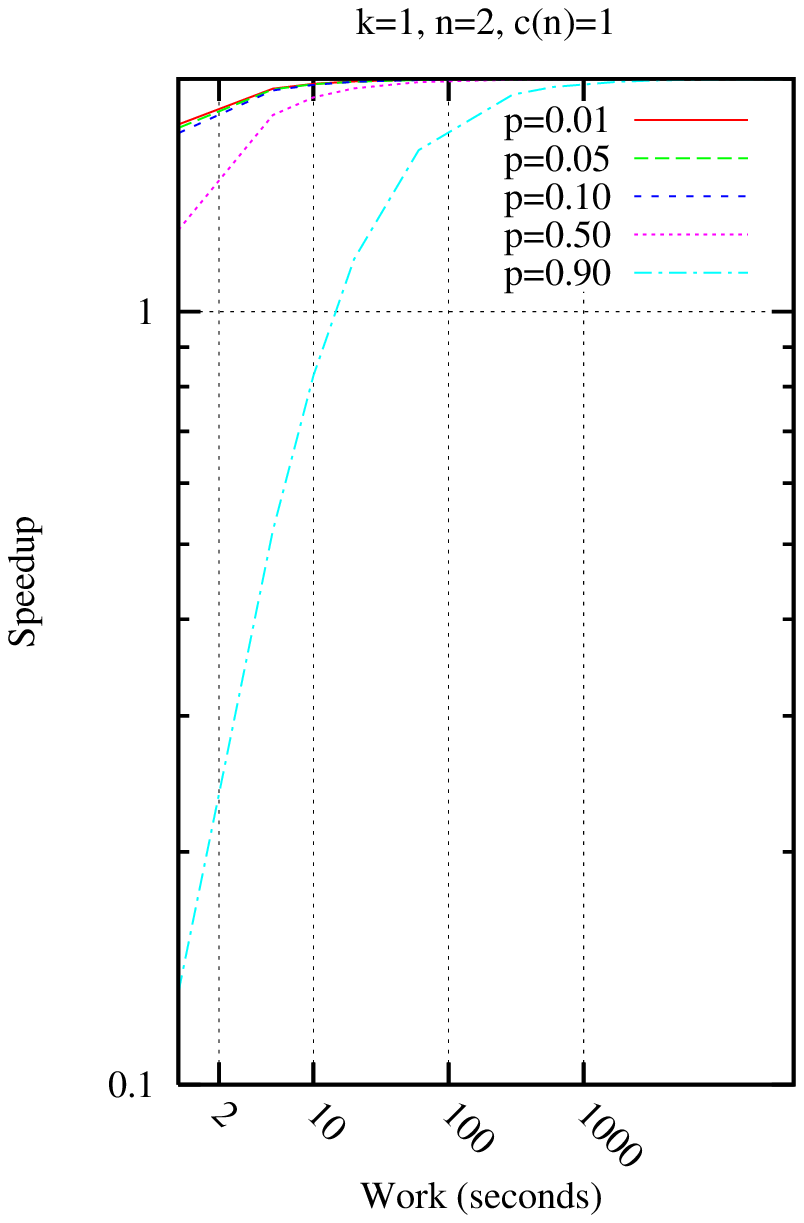}}
\subfigure[\label{SE_W_n2_logn}$c(n)=log_2(n)$]{\includegraphics[height=2.1in,width=2.1in]{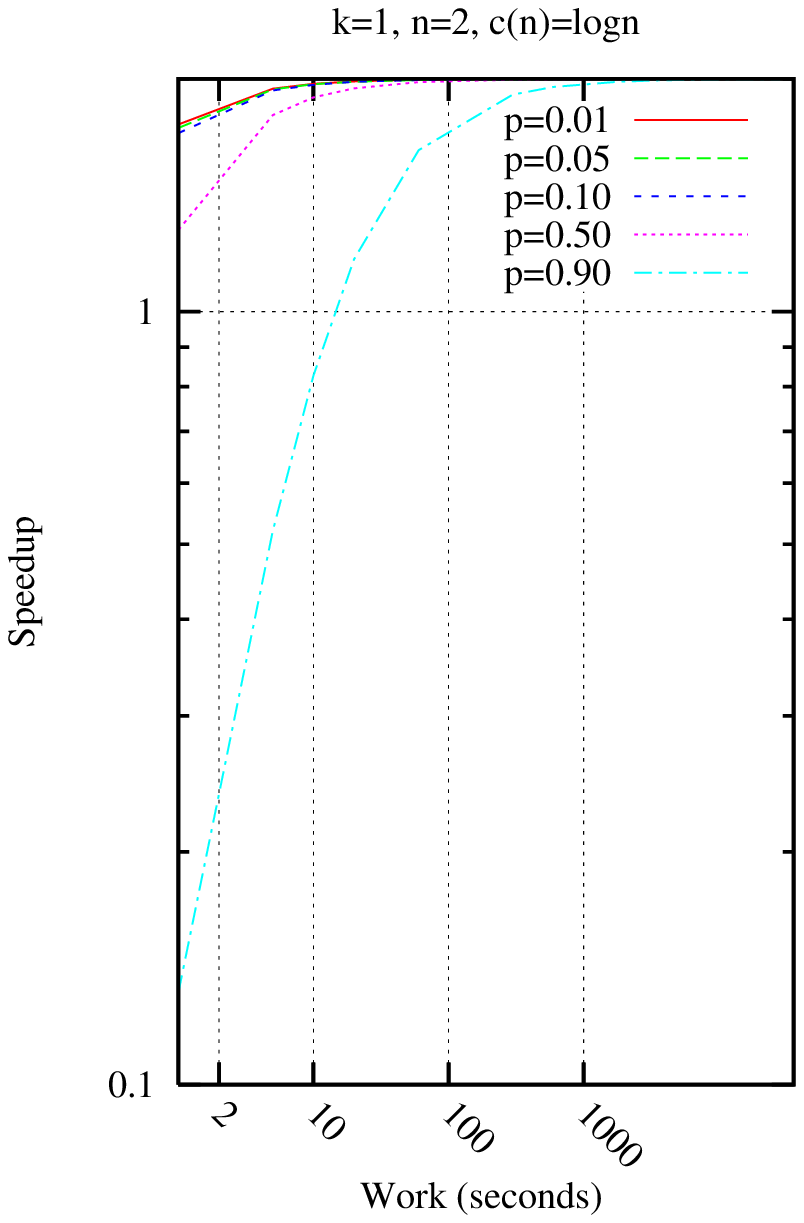}}
\subfigure[\label{SE_W_n2_log2n}$c(n)=log_2^2(n)$]{\includegraphics[height=2.1in,width=2.1in]{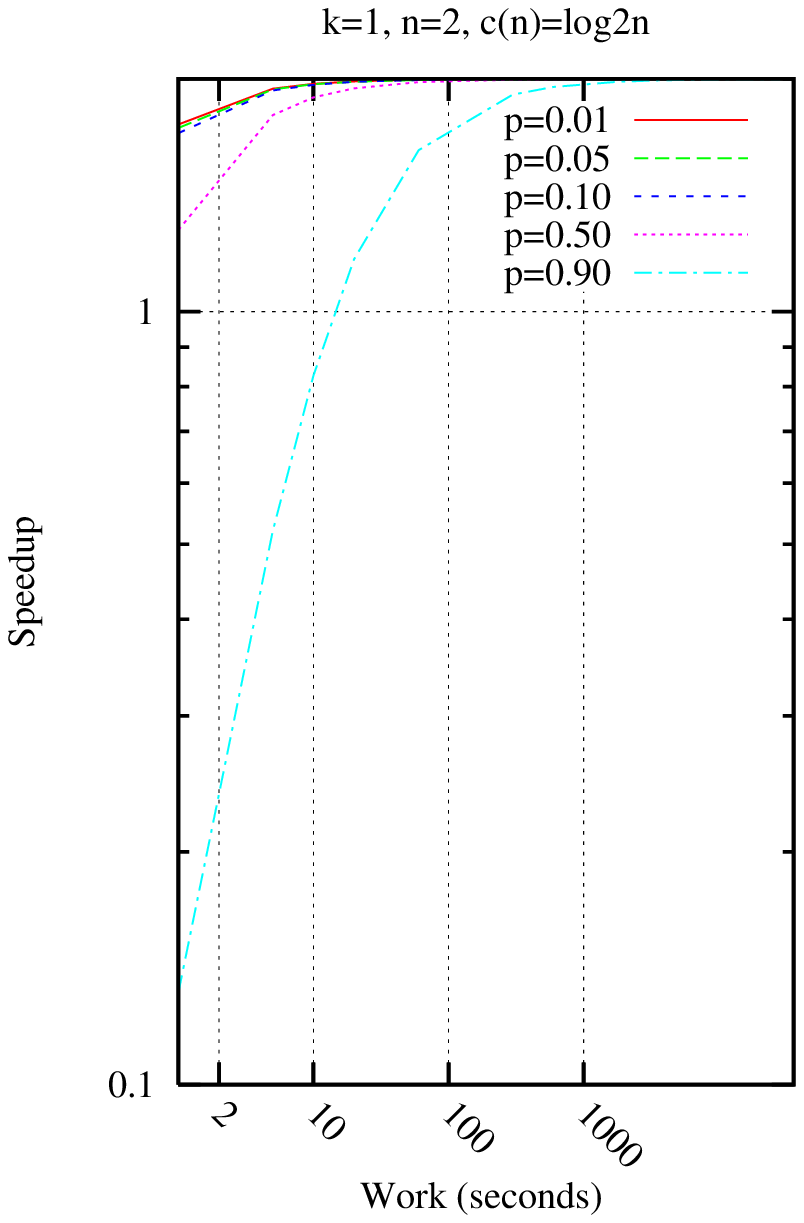}}}
\centerline{\subfigure[\label{SE_W_n2_n}$c(n)=n$]{\includegraphics[height=2.1in,width=2.1in]{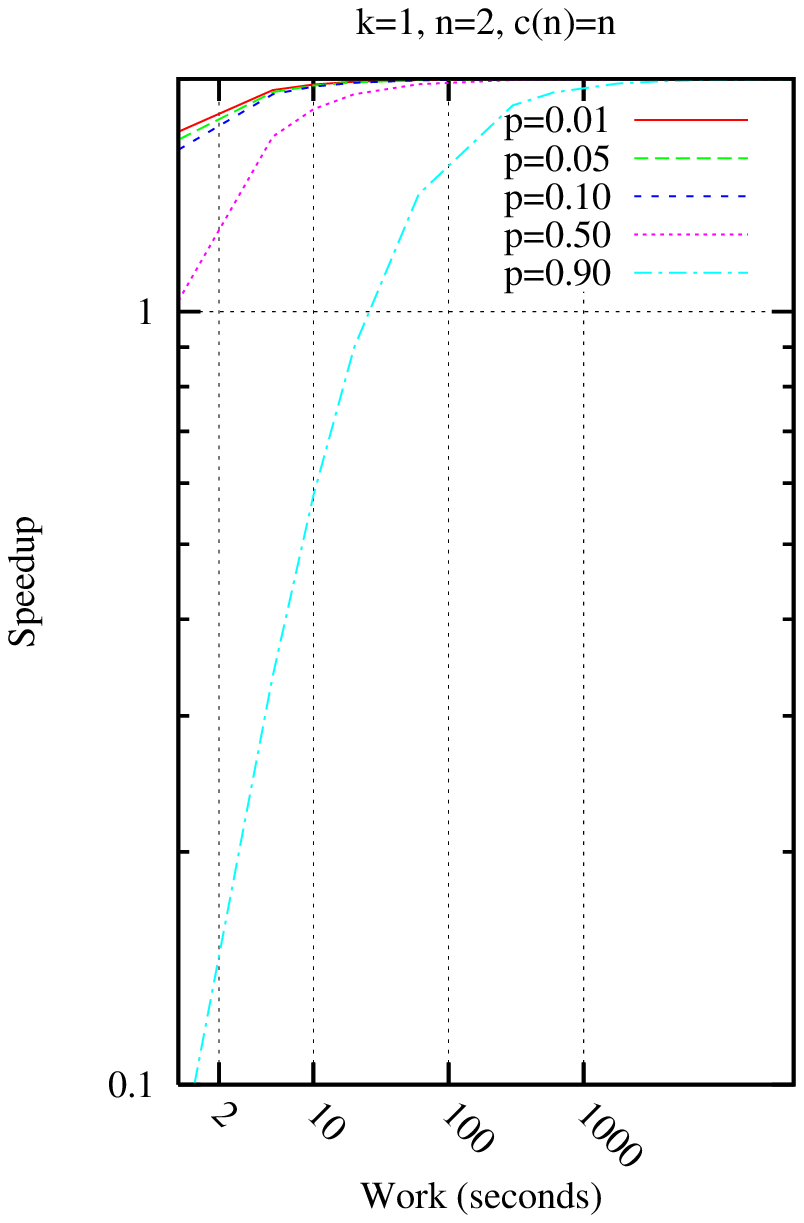}}
\subfigure[\label{SE_W_n2_nlogn}$c(n)=nlog_2(n)$]{\includegraphics[height=2.1in,width=2.1in]{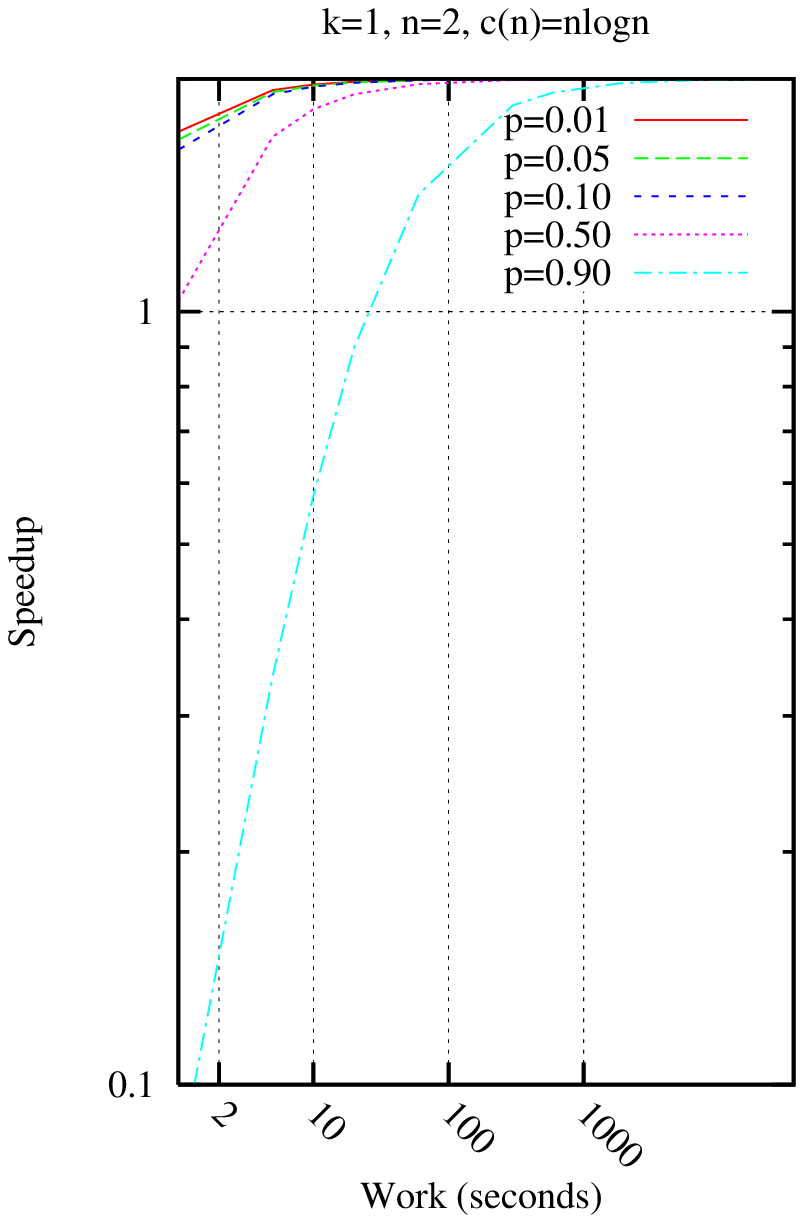}}
\subfigure[\label{SE_W_n2_n2}$c(n)=n^2$]{\includegraphics[height=2.1in,width=2.1in]{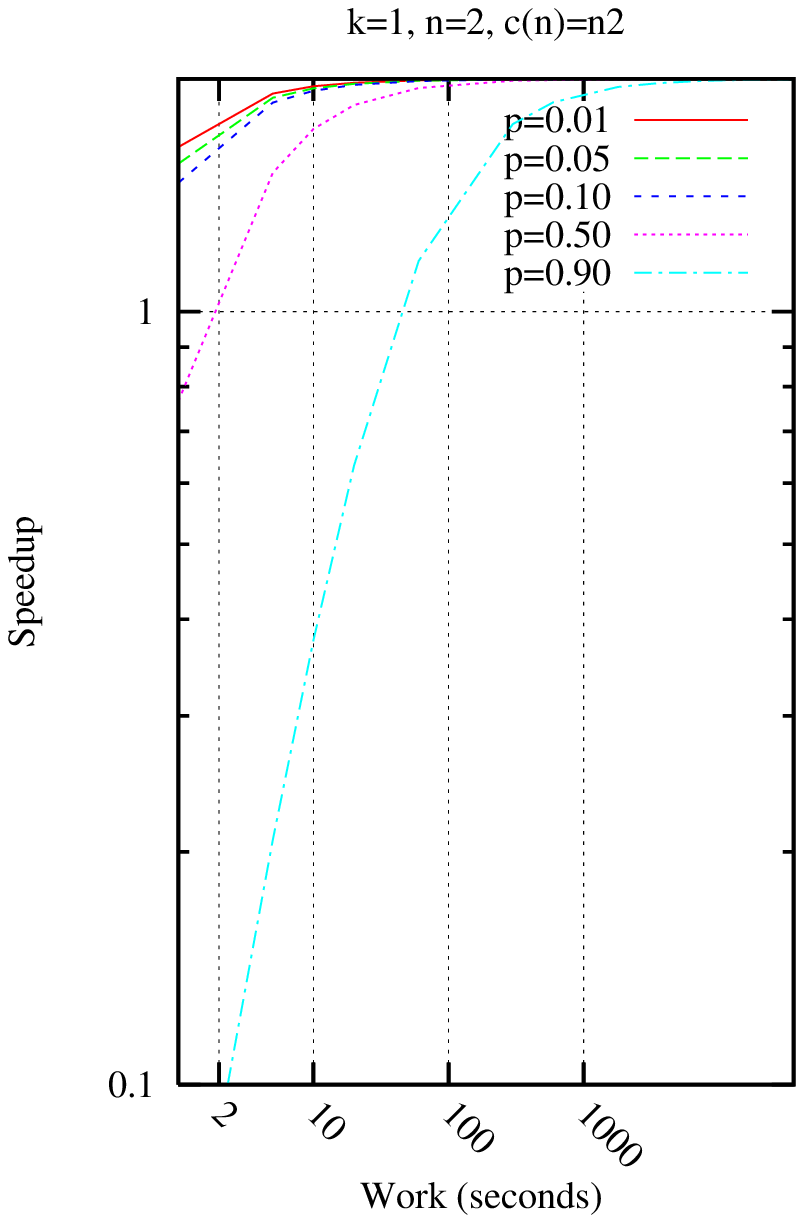}}}
\caption{\label{SE_W_n2}Graph depicts speedup for different work
sizes, for $n=2$, communication $c(n)$ at different packet loss
probabilities for the L-BSP model.}
\end{figure*}

\begin{figure*}[htbp]
\centerline{\subfigure[\label{SE_W_n131072_1}$c(n)=1$]{\includegraphics[height=2.1in,width=2.1in]{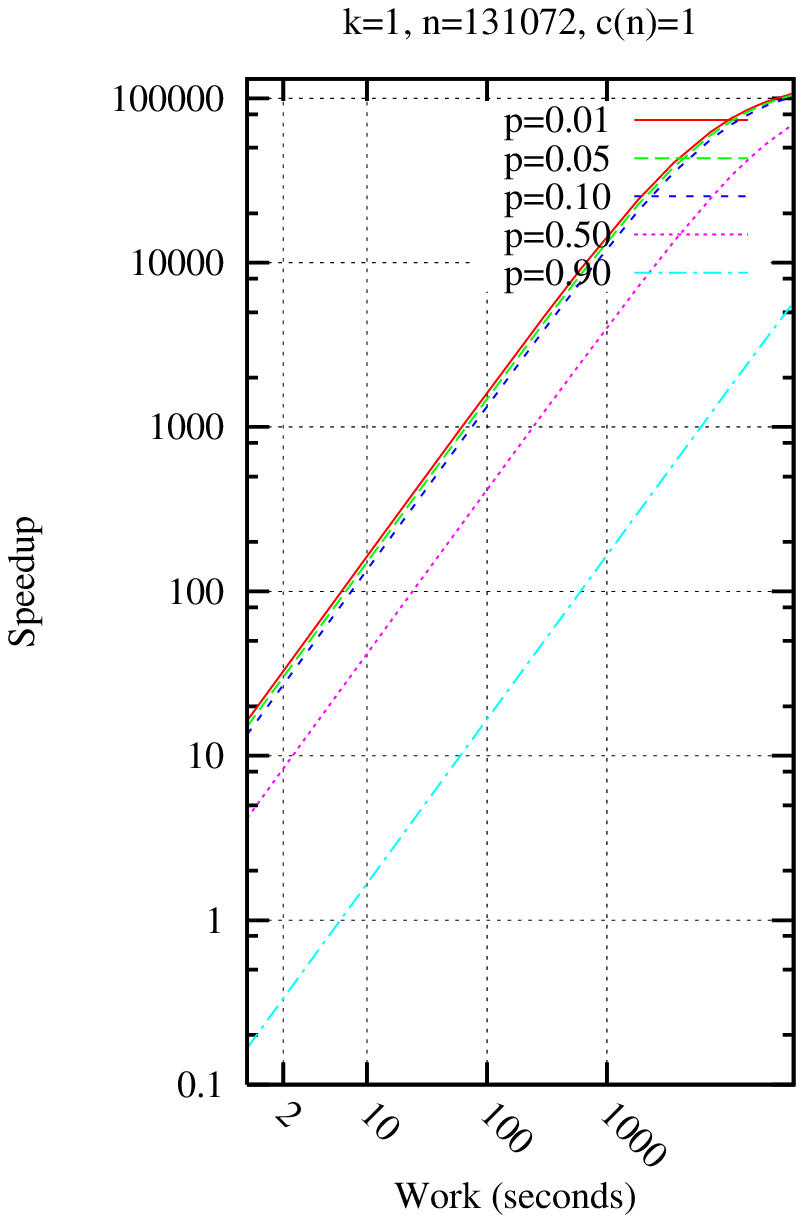}}
\subfigure[\label{SE_W_n131072_logn}$c(n)=log_2(n)$]{\includegraphics[height=2.1in,width=2.1in]{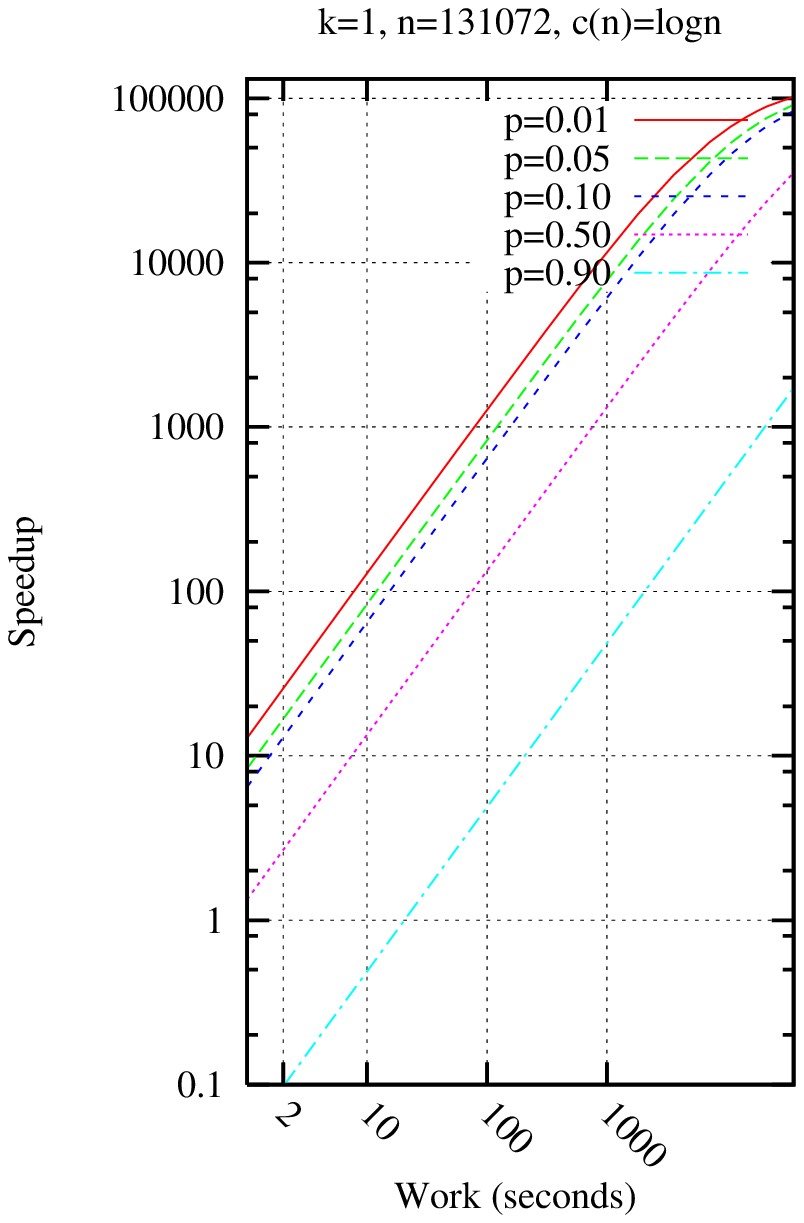}}
\subfigure[\label{SE_W_n131072_log2n}$c(n)=log_2^2(n)$]{\includegraphics[height=2.1in,width=2.1in]{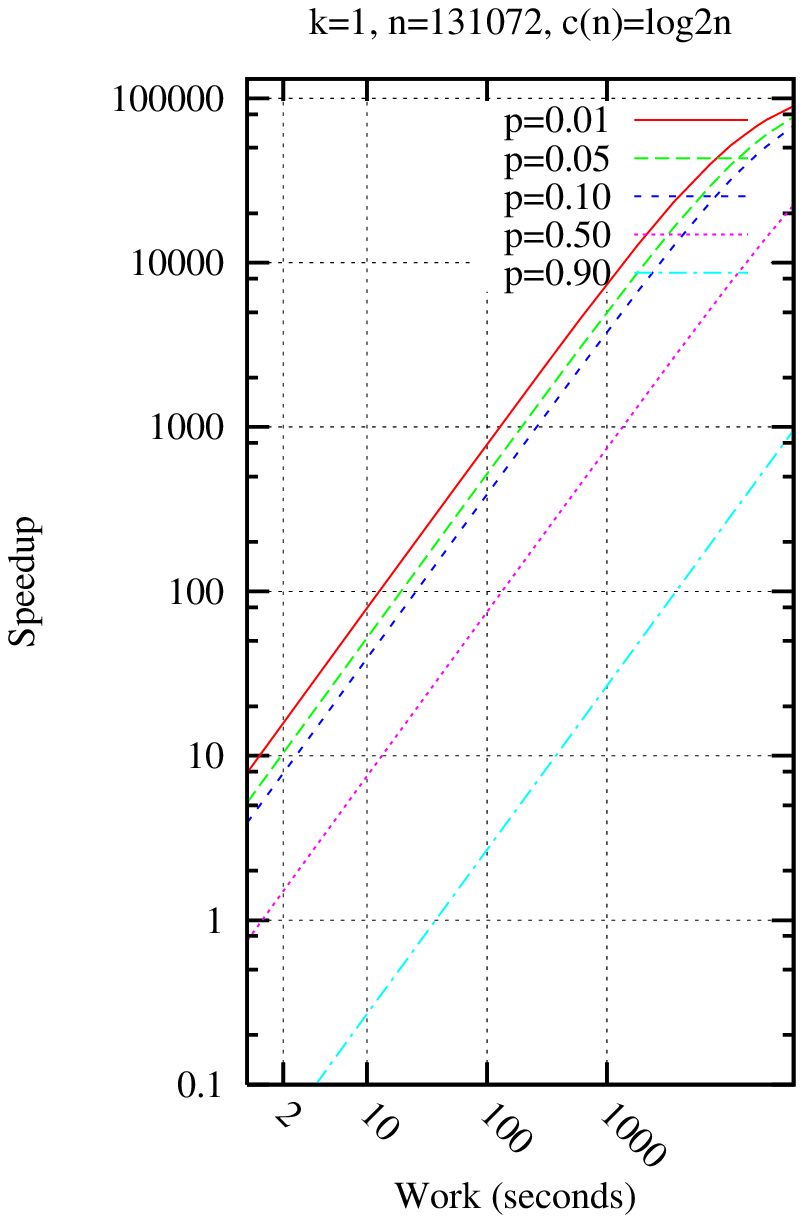}}}
\centerline{\subfigure[\label{SE_W_n131072_n}$c(n)=n$]{\includegraphics[height=2.1in,width=2.1in]{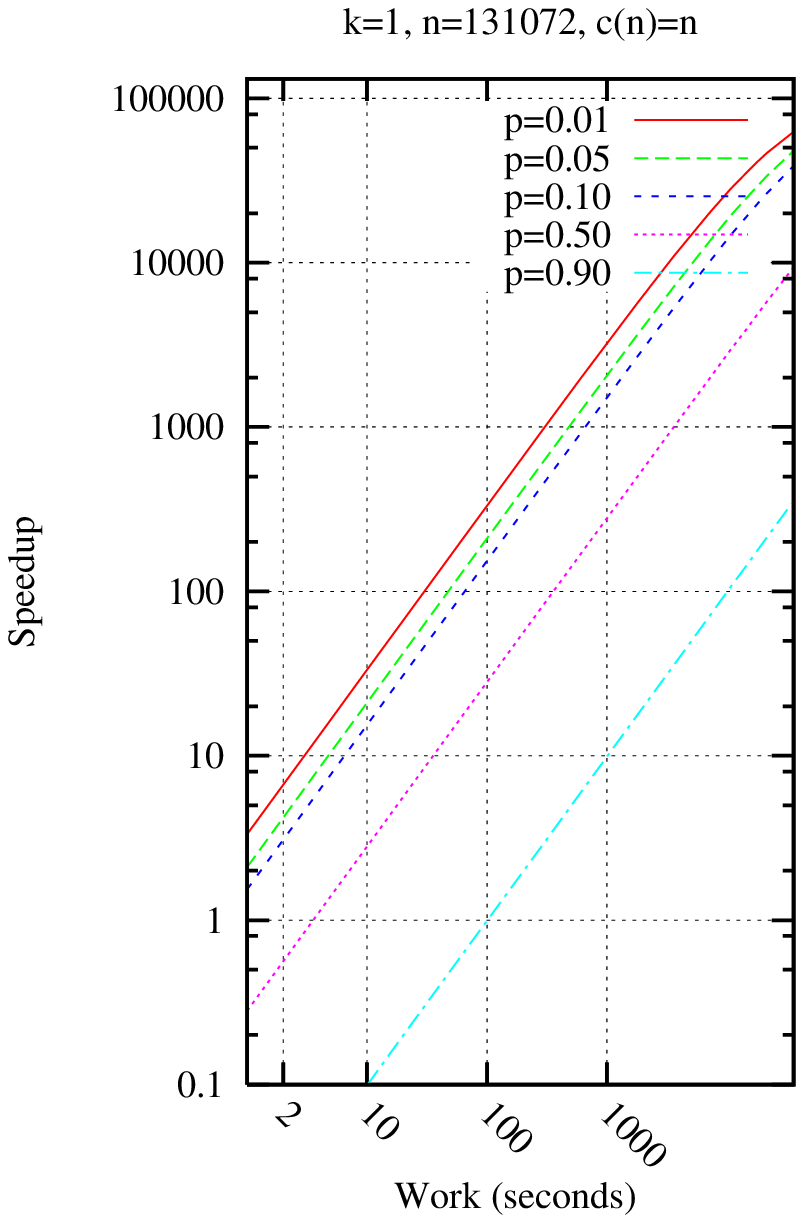}}
\subfigure[\label{SE_W_n131072_nlogn}$c(n)=nlog_2(n)$]{\includegraphics[height=2.1in,width=2.1in]{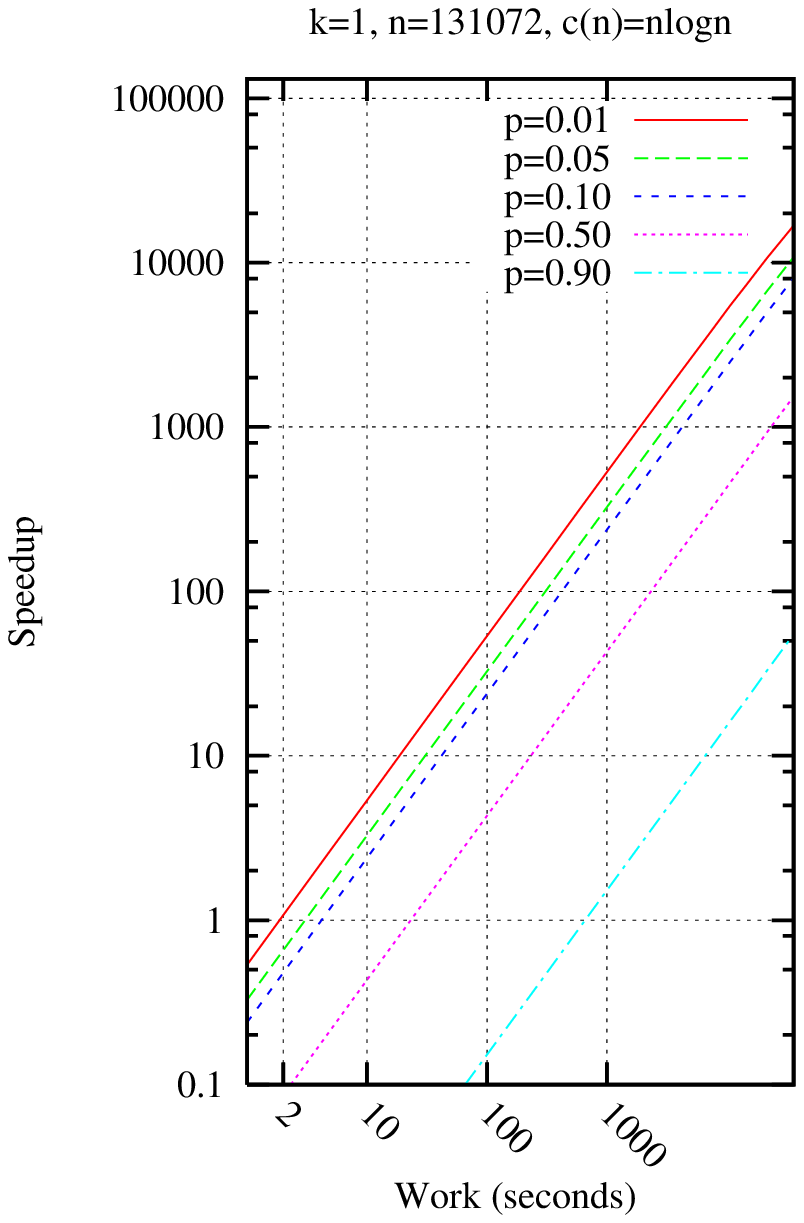}}
\subfigure[\label{SE_W_n131072_n2}$c(n)=n^2$]{\includegraphics[height=2.1in,width=2.1in]{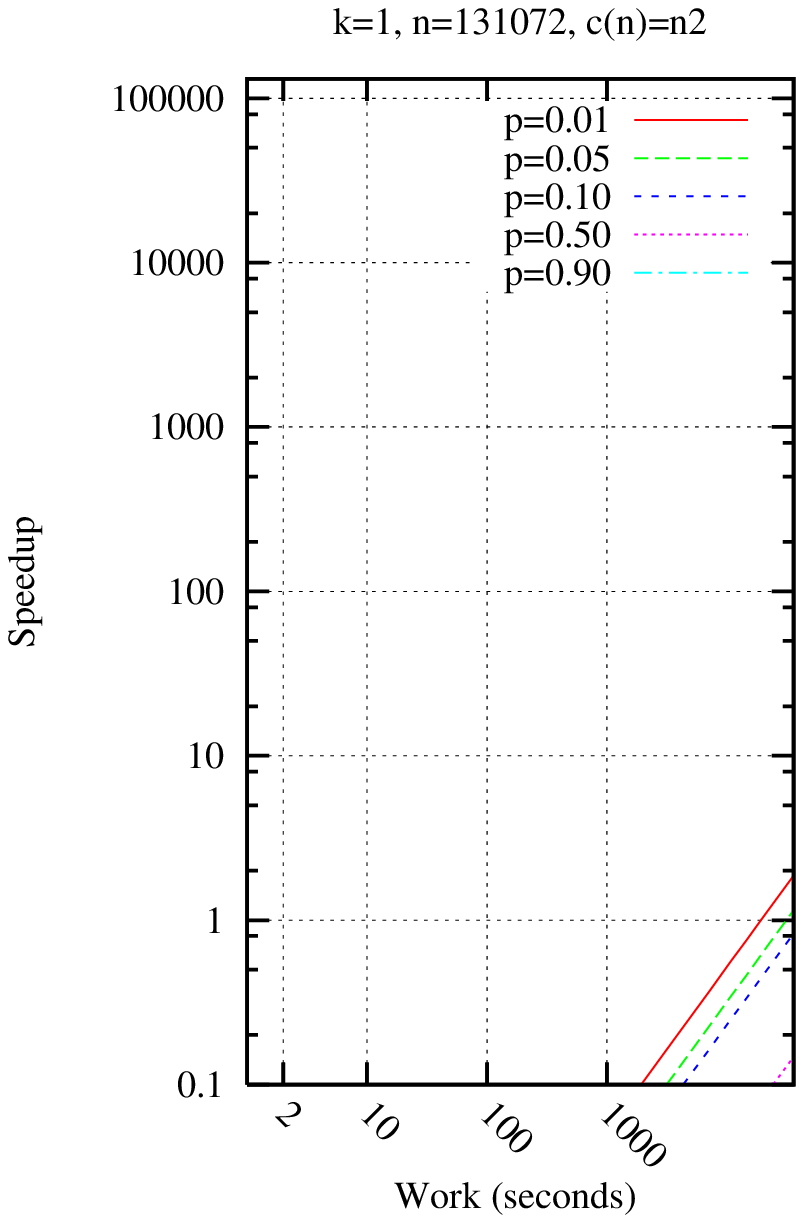}}}
\caption{\label{SE_W_n131072}Graph depicts speedup for different
work sizes, for $n=131072$, communication $c(n)$ at different
packet loss probabilities for the L-BSP model.}
\end{figure*}

\reffig{SE_W_n2} and \reffig{SE_W_n131072} shows predicted speedup
with different work loads, for different packet loss
probabilities. These graphs are for $n=2$ and $n=131072$ nodes
respectively when $k=1$. As the size of work increases on each
processor, speedup approaches the total number of processor used
for higher granularity.

\begin{comment}
\begin{figure*}[htbp]
\centerline{\subfigure[\label{SE_W_n256_1}$c(n)=1$]{\includegraphics[height=2.1in,width=2.1in]{gph_SE_W_k1_1_n256_comm.eps}}
\subfigure[\label{SE_W_n256_logn}$c(n)=log_2(n)$]{\includegraphics[height=2.1in,width=2.1in]{gph_SE_W_k1_logn_n256_comm.eps}}
\subfigure[\label{SE_W_n256_log2n}$c(n)=log_2^2(n)$]{\includegraphics[height=2.1in,width=2.1in]{gph_SE_W_k1_log2n_n256_comm.eps}}}
\centerline{\subfigure[\label{SE_W_n256_n}$c(n)=n$]{\includegraphics[height=2.1in,width=2.1in]{gph_SE_W_k1_n_n256_comm.eps}}
\subfigure[\label{SE_W_n256_nlogn}$c(n)=nlog_2(n)$]{\includegraphics[height=2.1in,width=2.1in]{gph_SE_W_k1_nlogn_n256_comm.eps}}
\subfigure[\label{SE_W_n256_n2}$c(n)=n^2$]{\includegraphics[height=2.1in,width=2.1in]{gph_SE_W_k1_n2_n256_comm.eps}}}
\caption{\label{SE_W_n256}Graph depicts speedup for different work
sizes, for $n=256$, communication $c(n)$ at different packet loss
probabilities for the L-BSP model.}
\end{figure*}
\end{comment}

\section{Adapting fundamental parallel algorithm}
In this section, we analyze some fundamental algorithms using the
L-BSP model, that provides the number of packet copies, $k$,
number of nodes, $n$, amount of work loads, $w$, to use depending
on packet loss probability, $p$, and communication complexity,
$c(n)$.

The L-BSP model in this paper considers sending data that fits
into a single packet. However, the maximum packet size in the
Internet Protocol Version 4(IPv4) is only $65$KB, to accommodate
large data we can: a) assume the usage of Internet Protocol
version 6 (IPv6) that provides maximum packet size of up to $4$GB.
b) Use multiple communication supersteps,$\gamma$, where
$\gamma=\big\lceil\frac{data\text{ }size}{packet \text{
}size}\big\rceil$.

Since it was not possible for us to obtain values for packet loss
probability, round-trip time and bandwidth using IPv6. This values
are extrapolated from our experiment using IPv4 on PlanetLab.

\begin{table*}[htbp]
\centering
  \caption{\label{algorithms}Approximate speedup of parallel algorithms for different parameter values using L-BSP model.}
\begin{tabular}{|c|c|c|c|c|}
 \hline
 Algorithm & Matrix multiplication,  & Bitonic & 2D-FFT & Laplace  Equation,\\
 & size ($N$x$N$) & Merge sort & & size ($N$x$N$)\\
 \hline
 Size, $N$ & $2^{15}$ & $2^{31}$ & $2^{34}$ & $2^{18}$\\
 \hline
 No. of processors, $n$ &$2^{16}$ &$2^{17}$ & $2^{15}$ & $2^{17}$\\
 \hline
 Size of message (bytes) & $2^{16}$&$2^{16}$ & $2^{8}$& $24$\\
 \hline
 Packet size, $p_{sz}$ & $2^{16}$& $2^{16}$& $2^{8}$& $24$\\
 \hline
 Packet copies, $k$ & $7$ & $6$ & $3$& $5$\\
 \hline
 Bandwidth,(MBytes/s) &$17.5$ & $17.5$& $17.07$& $24$\\
 \hline
 Packet loss probability, $p$ & $0.045$ &$0.045$ & $0.0005$& $0.0005$\\
 \hline
 $\frac{p_{sz}}{Bandwidth}, \alpha$ &$0.0037$ &$0.0037$&$0.000015$& $0.000001$ \\
 \hline
 Delay, $\beta$ & $0.069$ &$0.069$& $0.05$& $0.05$\\
 \hline
 Average No. of transmission, $\hat{\rho}^k$ & $1.025$& $1.002$ &$1.24$& $1.0$ \\
 \hline
 Sequential compute time (seconds), $w_s$ &$140765.34$ & $133.14$ &$5841.15$& $23364.44$\\
 \hline
 Communication cost (seconds) &$27.54$ &$28.18$& $7.35$& $1.7$\\
 \hline
 Total time in parallel (seconds) &$29.69$ &$28.194$ & $7.55$&$1.8783$\\
 \hline
 Communication complexity, $c(n)$& $\OO{n^\frac{3}{2}}$ &$\OO{n}$ & $\OO{n^2}$& $\OO{n}$\\
 \hline
 Average processor performance, (GFLOPS) & $0.5$&$0.5$ & $0.5$& $0.5$\\
 \hline
 Speedup, $S_E$ &$4740.89$ &$4.72$ & $773.4$& $12439.43$\\
\hline
 Efficiency & $0.072$ &$0.000036$ &$0.02$& $0.095$\\
\hline
\end{tabular}
\end{table*}

\subsection{Matrix multiplication (Direct implementation)}
\begin{figure}
\centering \tiny
\psfrag{sqtp}{$\sqrt{P}$}\psfrag{m}{$\frac{N}{\sqrt{P}}$}
\psfrag{b}{$A_{1,1}$}\psfrag{c}{$A_{1,2}$}\psfrag{d}{$A_{1,3}$}
\psfrag{e}{}\psfrag{f}{$B_{2,4}$}
\psfrag{g}{$B_{3,4}$}\psfrag{h}{$B_{4,4}$}
\includegraphics[height=2.0in,width=3.6in]{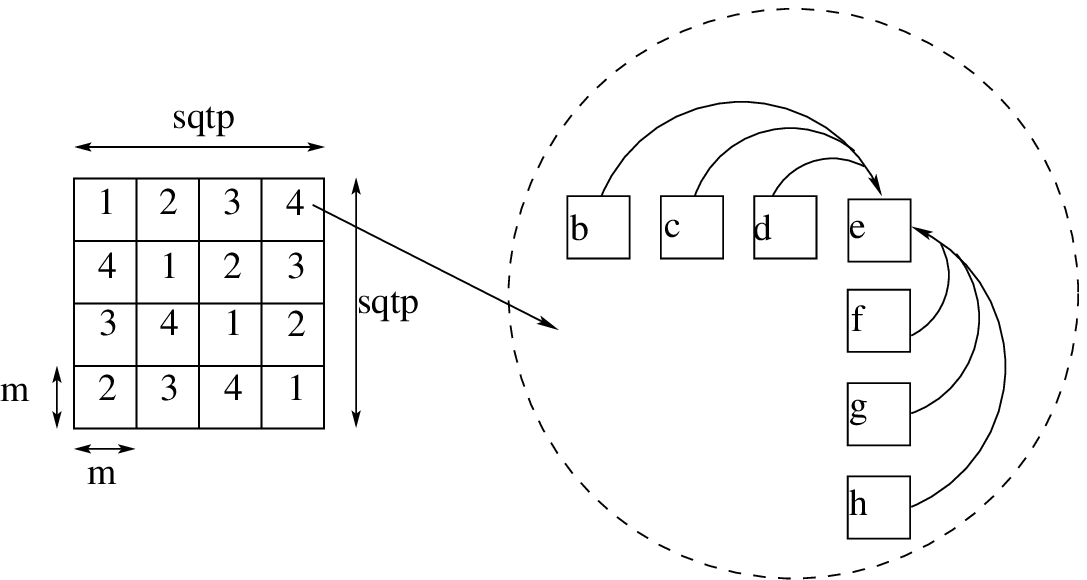}
\caption{\label{mat_mult} Parallel matrix multiplication on $P=16$
nodes.}
\end{figure}

Consider the product of two matrices $A$ and $B$ of size $N$x$N$,
producing matrix $C$ of size $N$x$N$, where $N \in 2^m$ and $m \in
\mathbb{Z}^+$. Each processor, $k$, with $k \in {1,2,\ldots,P}$
where $P$ is the total number of processor, has two submatrices of
size $\frac{N}{\sqrt{P}}$x$\frac{N}{\sqrt{P}}$, one from each
matrix (i.e. $A$ and $B$) and are indexed as $A_{i,j}$ and
$B_{i,j}$, where $i,j \in {1,2,\ldots,\sqrt{P}}$ denote rows and
columns respectively. Each submatrix contains $\frac{N^2}{P}$
elements. We assume each processor in the system have a submatrix
of $A$ and $B$ in them. Squares with same numbers in
\reffig{mat_mult} denotes submatrix $C_{i,j}$ that can be computed
concurrently. During the communication phase,
$c(P)=2(P^{\frac{3}{2}}-P)$ packets are injected into the network.
At the end of computation each processor has a portion of
submatrix $C$ in its possession. On a single processor the cost of
computing is $2N^3-N^2$. Using $P$ processor the cost of sending
submatrices from different processor to compute submatrix
$C_{i,j}$ as shown in \reffig{mat_mult} is
$2\gamma\hat{\rho}^k(2(\sqrt{P}-1)k\alpha+\beta)$ seconds. The
cost of computation is: $2\frac{N^3}{P}-\frac{N^2}{P}$ FLOPs with
the L-BSP model. Thus the expected speedup is:
\begin{equation}
S_E=\frac{w_s}{w_p+2\gamma\hat{\rho}^k(2(\sqrt{P}-1)k\alpha+\beta)}
\nonumber
\end{equation}
with, \\ $w_s=\frac{2N^3-N^2}{Average \text{ } FLOPS}$ and
$w_p=\frac{2\frac{N^3}{P}-\frac{N^2}{P}}{Average\text{ }FLOPS}$.
\\

Using this model, we analyzed achievable speedup for different
node sizes $P=2^s$ where $s=1,2,3,\ldots,17$ and for different
matrix dimensions, $N$x$N$ where $N=2^{11}, 2^{12}, 2^{13},
2^{14}, 2^{15}$. A best speedup of $4740.89$ is obtained when
$N=2^{15}$ and $P=2^{17}$, \reftable{algorithms} shows the
algorithm parameters for this speedup. The analysis shows that
matrix multiplication algorithm is very well suited for
parallelization on VLSG with our approach. Although the efficiency
is low at $0.072$, it is interesting to note that the problem is
solvable at almost $4741$ times faster using $2^{15}$ nodes
compared to on a single processor.

\subsection{Sorting (Bitonic mergesort)}

Here, we analyze the complexity of Batcher's bitonic sort
algorithm~\cite{Batcher68}. Assuming each processor in the system
has $N$ unsorted keys, rearrange them so that every key in
processor $i$ is less than or equal to every key in processor
$i+1$ where $1\le i< P$. First, each processor sorts its
$\frac{N}{P}$ keys locally (either ascending or descending order,
for obtaining bitonic sequence). Then this algorithm does
$log_2(P)$ merge stages as shown in \reffig{bitonic}, where stage
$S$ $(1\le S\le log_2(P))$ has $S$ merge steps. In merge step $j$
$(1\le j \le S)$ of stage $S$, each processor $i$ sends the list
of $\frac{N}{P}$ keys in its possession to the processor $x$ ($x$
is obtained by complementing the $j$th bit of $i$). Thereafter,
every processor does the merging and keeps either the first half
or the second half of the merged list. At the end of sorting, each
processor $i$ will have $\frac{N}{P}$ sorted keys that are less or
equal to every key in processor $i+1$.~\cite{Juurlink98} A total
of $\frac{log_2(P)(log_2(P)+1)}{2}$ steps are required in this
algorithm. In each step, a total of $c(P)=P$, UDP packets are
transmitted. The computational cost of sorting an unsorted
sequence of size $N$ and merging them in parallel is
$\frac{N}{P}log_2\big(\frac{N}{P}\big)+\frac{log_2(P)(log_2(P)+1)}{2}\big(\frac{2N}{P}-1\big)$
FLOPs, and the total communication cost is
$\gamma(log_2(P)(log_2(P) + 1))(k\alpha+\beta)\hat{\rho}^k$
seconds. The expected speedup can be calculated using the L-BSP
model as:
\begin{equation}
S_E=\frac{w_s}{w_p+\gamma
log_2(P)(log_2(P)+1)(k\alpha+\beta)\hat{\rho}^k} \nonumber
\end{equation}
with, $w_s=\frac{Nlog_2N}{Average \text{ } FLOPS}$ and
$w_p=\frac{\frac{N}{P}log_2\big(\frac{N}{P}\big)+log_2(P)(log_2(P)+1)\big(\frac{N}{P}-\frac{1}{2}\big)}{Average\text{
} FLOPS}$.

\begin{figure}[htbp]
\centering \tiny
\psfrag{1}{$1$}\psfrag{2}{$2$}\psfrag{3}{$3$}\psfrag{4}{$4$}
\psfrag{P}{$P$}\psfrag{P-3}{$P-3$}\psfrag{P-2}{$P-2$}\psfrag{P-1}{$P-1$}
\psfrag{logp}{$log_2(P)$}\psfrag{S=1}{$S=1$}\psfrag{S=2}{$S=2$}
\psfrag{S=logp}{$S=log_2(P)$}\psfrag{j=1}{$j=1$}
\psfrag{j=2}{$j=2$} \psfrag{j=logp-1}{$j=log_2(P)-1$}
\psfrag{j=logp}{$j=log_2(P)$} \psfrag{p-4}{$P-4$}
\psfrag{p-3}{$P-3$} \psfrag{p-2}{$P-2$}
\psfrag{p-1}{$P-1$}\psfrag{w(w+1)/2}{$\frac{log_2(P)(log_2(P)+1)}{2}$}
\includegraphics[height=2in,width=3.6in]{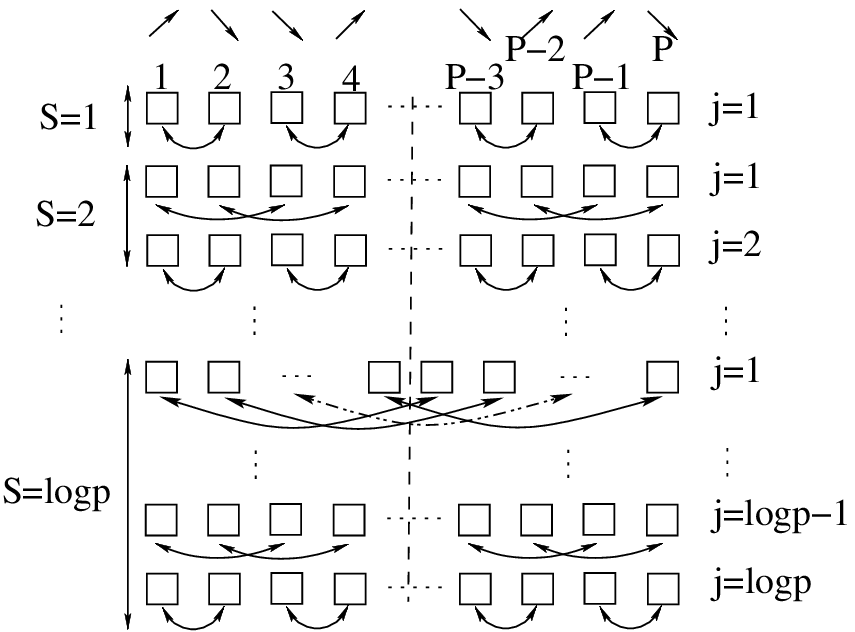}
\caption{\label{bitonic} Parallel bitonic mergesort for $P$
nodes.}
\end{figure}

With this model, achievable speedup for different sizes of data,
$N=2^{20}, 2^{24}, 2^{28}, 2^{29}, 2^{30} ,2^{31}$ and different
number of nodes $P=2^s$ where $s=1,2,3,\ldots,17$ were analyzed.
The best speedup of $4.722$ was obtained when $P=2^{17}$ and
$N=2^{31}$, \reftable{algorithms} shows the algorithm parameters
for this speedup. Although, efficiency is very low for this
algorithm it is interesting to observe that some speedup can still
be obtained on VLSG with our approach.

\subsection{2D Fast Fourier Transform (FFT) transpose method (FFT-TM)}
Fourier transform plays an important role in many scientific and
technical applications. A fast Fourier transform (FFT) has a RAM
complexity of $\OO{NlogN}$. The FFT can be parallelized using the
2D FFT-TM algorithm. The 2D FFT-TM algorithm can be viewed as
computing multiple 1D FFTs in each direction using a fast FFT
library (e.g. FFTW) and it has a couple of all-to-all
communication for inter-processor data distribution. During this
communication, each node will send a portion of its $\frac{N}{P}$
data (i.e. $\frac{N}{P^2}$ data) to $P-1$ processors. The received
data (complex numbers with datum size of $16$ bytes) is then
re-arranged to complete the transpose process. We assume the cost
of rearranging the data to be insignificant. In our analysis, all
the processor has $\frac{N}{P}$ data in their node initially and
only one UDP data packet is required to send $\frac{N}{P^2}$
portion of its data to another node. A total of $c(P)=P(P-1)$ UDP
data packets of size $\frac{Nb}{P^2}$, where $b$ is the data size,
are transmitted during the all-to-all data distribution. The total
parallel computation cost for this algorithm is given by
$10\frac{N}{P}log(\frac{N}{P})$ FLOPs and the communication cost
is $4\gamma\hat{\rho}^k\big(k\alpha (P-1)+\beta\big)$ seconds.
Following, the expected speedup can be calculated using L-BSP
model as:
\begin{equation}
S_E=\frac{w_s}{w_p+4\gamma\hat{\rho}^k\big(k\alpha(P-1)+\beta\big)}
\nonumber
\end{equation}
with, $w_s=\frac{5Nlog(N)}{Average\text{ }FLOPS}$ and
$w_p=\frac{10\frac{N}{P}log\big(\frac{N}{P}\big)}{Average\text{ }
FLOPS}$.

Using this model, speedup for different node sizes $P=2^s$ where
$s=1,2,3,\ldots,15$ and for different data sizes, $N$ where
$N=2^{30}, 2^{32}, 2^{34}, 2^{36}, 2^{38}$ were analyzed. A best
speedup of $773.4$ with efficiency at $0.02$ is obtained when
$N=2^{34}$ and $P=2^{15}$. \reftable{algorithms} shows the
algorithm parameters for this speedup. The analysis shows that
2D-FFT algorithm maybe suitable for parallelization on VLSG with
our approach for very large data size.

\subsection{Partial differential equation: The Laplace's Equation}
Partial differential equation is used in many different fields of
computational science to model real world phenomena. One of the
fundamental partial differential equation is the Laplace's
equation:
\begin{align}
&\frac{\partial^2f}{\partial x^2}+\frac{\partial^2f}{\partial
y^2}=0, \nonumber \\
& f(0,y)=U_1(y), f(l,y)=U_2(y), \nonumber \\
& f(x,0)=U_3(x), f(x,l)=U_4(x), \nonumber \\
& 0<x<l, 0<y<l. \nonumber
\end{align}
The solution $f(x,y)$, for this equation with given boundary
conditions can be found using the finite difference method. First
the Laplace equation must be discretized and the resulting system
of linear equation is then solved. In matrix-vector form, this
system of linear equation has a sparse matrix with $5$ (nonzero
elements) diagonals. In this paper, we analyze the Jacobi method
used to solve this system.  We can represent the function $f(x,y)$
by its values at discrete set of uniformly-spaced network of mesh
points, $x_i=i\Delta x$ and $y_j=j\Delta y$ for $i=0,1,\ldots ,m$
and $j=0,1,\ldots ,q$ with $\Delta x =\frac{l}{m}$ and $\Delta y =
\frac{l}{q}$. The grid size $h$ of $x$-dimension and $y$-dimension
are chosen to be equal, $h=\Delta x = \Delta y$ to simplify our
analysis. The function $f(x,y)$ at any point $(x_i,y_j)$ is
represented by $f_{i,j}$. The finite difference representation for
Laplace equation is,
\begin{equation}
\frac{f_{i+1,j}-2f_{i,j}+f_{i-1,j}}{h^2}+\frac{f_{i,j+1}-2f_{i,j}+f_{i,j-1}}{h^2}=0.
\nonumber
\end{equation}
and can be rearranged as,
\begin{equation}
f^{k+1}_{i,j}= \frac{1}{4}
\big[f^{k}_{i+1,j}+f^{k}_{i-1,j}+f^{k}_{i,j+1}+f^{k}_{i,j-1}\big].
\nonumber
\end{equation}
where $f^{k+1}$ is the value obtained from $k+1$th iteration and
$f^k$ is the value obtained from the $k$th iteration. The
$(i,j)$th point is computed from the $ij$th (product of $i$ and
$j$) equation. This results in a system of linear equation with
$(m-1)^2$ equations and $(m-1)^2$ unknowns. Jacobi method can be
used to solve this equation and is described as:
\begin{equation}
x_i^k=\frac{1}{a_{i,i}}\big[b_i-\sum_{j\neq
i}a_{i,j}x_j^{k-1}\big] \nonumber
\end{equation}

For a pentadiagonal system with $\frac{(m-1)^2}{P}>5$, ($P$ is the
number of processors used) each node is required to exchange at
most $3$ newly calculated values of unknowns between neighboring
nodes. Thus, $c(P)=2(P-1)$ packets of size $3b$ bytes, where $b$,
is the data size in bytes, are injected into the network at any
one communication. For a diagonally dominant matrix, which is the
case for Laplace equation, the Jacobi method will converge to ``a
good solution" in $log_2P$ steps, however this depends on the
initial value used and the convergence rate. In our analysis, we
take $log_2P$ as the number of rounds required for convergence.

The total parallel computation cost for this algorithm is
$2dlog_2P\big(\frac{(m-1)^2}{P}\big)$ FLOPs, where $d$ is the
number of diagonals in the matrix ($d=5$ for a pentadiagonal
system). The total communication time is
$2log_2P\hat{\rho}^k\big(\alpha k \frac{2(P-1)}{P}+\beta\big)$
seconds. Following, the expected speedup can be calculated using
L-BSP model as:
\begin{equation}
S_E=\frac{w_s}{w_p+2\hat{\rho}^klog_2P\big(k\alpha\frac{2(P-1)}{P}+\beta\big)}
\nonumber
\end{equation}
with, $w_s=\frac{2dlog_2P(m-1)^2}{Average\text{ }FLOPS}$ and
$w_p=\frac{2dlog_2P\big(\frac{(m-1)^2}{P}\big)}{Average\text{ }
FLOPS}$.
\begin{center}
\begin{figure}[htbp]
\centering \tiny
\psfrag{m/P}{$\frac{(m-1)^2}{P}$}\psfrag{m-1}{${(m-1)}^2$}
\psfrag{exchg}{Exchange}
%\psfrag{send to every other node}{send
%toevery other node}
\includegraphics[height=2in,width=3.2in]{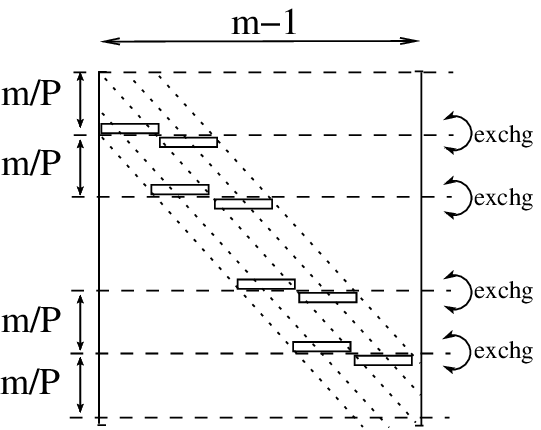}
\caption{\label{bitonic} Each node computes $\frac{(m-1)^2}{P}$
points and exchanges at most $3$ newly computed values in a
pentadiagonal system.}
\end{figure}
\end{center}
With this model, achievable speedup for different dimension
$m$x$m$, where $m=2^{14}, 2^{15}, 2^{16}, 2^{17} ,2^{18}$ and
different number of nodes $P=2^s$ where $s=1,2,3,\ldots,17$ were
analyzed. The best speedup of $12439.43$ was obtained when
$P=2^{17}$ and $m=2^{18}$, \reftable{algorithms} shows the
algorithm parameters for this speedup. Although, efficiency is low
at $0.095$ for this algorithm, it is interesting to observe that
reasonable speedup can still be obtained on VLSG with our
approach.

\subsection{Broadcast}
The broadcast operation is a fundamental primitive in many
parallel applications. A broadcast is a communication pattern
where a source node sends messages to all other processors in the
system. Two commonly used algorithms are the binomial tree for
short messages and the one proposed by Van de
Geijn~\cite{Shroff99} for long messages. In the binomial tree
algorithm, the root node $P_0$ sends data to node
$P_{0+\frac{P}{2}}$. These nodes then act as the new roots within
their own subtrees and recursively distributes the messages. This
communication takes a total of $\lceil logP \rceil$ steps, in the
following analysis, we consider messages that fit into a single
packet. In step $logP$, $c(P)=logP$ packets are communicated.
Thus, the total cost of communication in L-BSP model can be
simplified as:
\begin{equation}
t_{bcast}= \bigg[\frac{k\alpha}{P}(1-2^{\lceil logP\rceil
-1})+\beta \lceil logP\rceil\bigg]\hat{\rho}^k \nonumber
\end{equation}
%For a $d-ary$ tree, the height of the tree is $h=\lceil log_d((d-1)P+1)-1\rceil$.
\subsection{All-gather}
The all-gather is an operation where data from different
processors are gathered on all processors. Three different
algorithms can be used for this operation: the ring method,
recursive doubling and Bruck algorithm. In the ring method each
processor $i$ sends its portion of data to processor $i+1$ and
receives data from processor $i-1$ (with a wrap around). In the
next step, each processor $i$ forwards to processor $i+1$ the data
it received from $i-1$ in the previous step. These steps are
repeated for $P-1$ times, where $P$ is the number of processor. If
$N$ is the total number of data to be gathered on each processor,
then at every step each processor sends $\frac{N}{P}$ data from
$P-1$ other processors.\cite{Thakur05} Here we consider data size
that fit into a single packet. Thus, the total cost of
communication for ring method in the L-BSP model is,
$t_{allgather}=\big[k\alpha+\beta\big](P-1)\hat{\rho}^k$.

\section{Related work}
Many computational models have been developed for parallel
architecture, each of them try to reflect the behavior of parallel
algorithms running on parallel architecture. As stated by Maggs et
al~\cite{Maggs95}, models should balance between simplicity with
accuracy, abstraction with practicality, and descriptivity with
prescriptivity. Early models such as PRAM~\cite{8} and its
variants that emphasize on PRAMs weakness (e.g. Phase
PRAM~\cite{15}, APRAM~\cite{21}, LPRAM~\cite{12}, and
BPRAM~\cite{16}) have been developed for parallel architectures.
Other models such as Postal model~\cite{24}, BSP~\cite{valiant90}
and LogP~\cite{13} that considers communication costs such as
network latency and bandwidth were developed to better reflect
behavior of parallel algorithms. Variants of BSP such as
D--BSP~\cite{30,29}, E--BSP~\cite{23} and CGM~\cite{26,27,28,25}
and models that take memory hierarchy into consideration such as
Parallel Hierarchical Memory model (P-HMM)~\cite{37}, LogP-HMM and
LogP-UMH~\cite{38} have also been developed. However, there is no
single model that has become a standard choice for the parallel
computing community. This is due to the heterogeneity in
communication topology and architecture. Computational model for
grid platform is still at its infancy and not many models have
been developed for this platform yet. The two well known
computational models for grid are the k-Heterogeneous Bulk
Synchronous Parallel (HBSP$^k$)~\cite{ williams00},
DynamicBSP~\cite{Martin04} and BSPGRID~\cite{Vasilev03}. Although
some of these models do consider communication and network
latency, they, however, do not consider packet loss as a
fundamental parameter in their models. This could be because most
of these models assume the usage of TCP protocol for internode
communication purposes and other factors in TCP that far outweighs
the impact of packet losses. In our approach, we used UDP as the
communication protocol. It is well known that packet loss and
congestion control mechanism (such as rate based congestion
control) are the main contributor to UDPs performance. In this
paper we concentrate our attention on the impact of packet loss on
performance and the usage of multiple packet copies to improve
performance.

\section{Conclusion}
Providing an accurate model to reflect the behavior of sequential
program running on a single computer is difficult due to current
technologies in computer architecture. On grids, it becomes even
harder as there are many more factors that influence the behavior
of computing resources and network.

Experiments that run parallel programs on PlanetLab indicates that
the communication phase between nodes on a WAN is the main
bottleneck effecting performance of parallel programs, even when
computing resources are not highly loaded by other jobs. Thus, it
is very necessary to utilize the fastest available protocol (UDP)
to execute parallel programs on grids. The weakness of UDP can be
remedied by using a light-weight mechanism for reliability to
enhance achievable speedup.

In this paper, a new model based on the BSP model that considers
packet loss probability as a fundamental parameter is introduced.
We measured average packet loss, round trip time, and bandwidth
for UDP between pairs of nodes within PlanetLab to better
understand the dynamics of WAN. This information is then used in
our model to derive the optimal number of packet copies to use in
order to maximize the speedup of parallel programs. The effect of
packet loss on performance of parallel programs is shown.

We also analyzed a few fundamental algorithms using the L-BSP
model. Although the efficiency is very low in some cases, the
result shows that it is possible to obtain some speedup when large
number of nodes are used. It is also important to note that the
result do not incorporate the effect of memory hierarchy.

In future work, other features such as replication of parallel
program for fault tolerance and reliability are being considered.
We intend to evaluate the performance of parallel program based on
L-BSP model and detailed packet loss model~\cite{padyee98} for
TCP.

\section*{Acknowledgment}
% optional entry into table of contents (if used)
%\addcontentsline{toc}{section}{Acknowledgment}
Elankovan Sundararajan would like to thank The National University
of Malaysia and The University of Melbourne for providing
financial assistance.

\bibliographystyle{IEEEtran}%latex8}
\bibliography{IEEEabrv,UDP}

\begin{thebibliography}{10}
\providecommand{\url}[1]{#1}
\csname url@rmstyle\endcsname
\providecommand{\newblock}{\relax}
\providecommand{\bibinfo}[2]{#2}
\providecommand\BIBentrySTDinterwordspacing{\spaceskip=0pt\relax}
\providecommand\BIBentryALTinterwordstretchfactor{4}
\providecommand\BIBentryALTinterwordspacing{\spaceskip=\fontdimen2\font plus
\BIBentryALTinterwordstretchfactor\fontdimen3\font minus
  \fontdimen4\font\relax}
\providecommand\BIBforeignlanguage[2]{{%
\expandafter\ifx\csname l@#1\endcsname\relax
\typeout{** WARNING: IEEEtran.bst: No hyphenation pattern has been}%
\typeout{** loaded for the language `#1'. Using the pattern for}%
\typeout{** the default language instead.}%
\else
\language=\csname l@#1\endcsname
\fi
#2}}

\bibitem{bouteiller03coordinated}
A.~Bouteiller, P.~Lemarinier, G.~Krawezik, and F.~Cappello, ``Coordinated
  checkpoint versus message log for fault tolerant {MPI},'' in \emph{Cluster
  2003}, 2003.

\bibitem{Postel81}
J.~Postel, ``{RFC 793:Transmission Control Protocol},'' September 1981.

\bibitem{Postel80}
------, ``{RFC 768:User Datagram Protocol},'' August 1980.

\bibitem{Irwin92}
B.~Irwin and M.~Mathis, ``{Web100: Facilitating High-Performance Network Use.
  White Paper for the Internet2 End-to-End Performance Initiative.}''

\bibitem{Jacobson92}
V.~Jacobson, R.~Braden, and D.~Borman, ``{RFC 1323:TCP Extensions for High
  Performance },'' May 1992.

\bibitem{Yuhong04}
Y.~Gu and R.~Grossman, ``{SABUL: A Transport Protocol for Grid Computing},''
  \emph{Journal of Grid Computing}, vol.~1, no.~4, pp. 377--386, 2004.

\bibitem{Feng00}
W.~Feng and P.~Tinnakornsrisuphap, ``{The failure of TCP in high-performance
  computational grids},'' in \emph{Supercomputing '00: Proceedings of the 2000
  ACM/IEEE conference on Supercomputing (CDROM)}.\hskip 1em plus 0.5em minus
  0.4em\relax Washington, DC, USA: IEEE Computer Society, 2000.

\bibitem{Liu02}
P.~Liu, M.~Meng, Y.~Xiufen, and J.~Gu, ``{An UDP-based protocol for Internet
  robots},'' \emph{Proceedings of the 4th World Congress on Intelligent Control
  and Automation}, vol.~1, pp. 59--65, 2002.

\bibitem{Mark}
M.~Meiss, ``{Tsunami: A High-Speed Rate-Controlled Protocol for File
  Transfer},'' http://steinbeck.ucs.indiana.edu/~mmeiss/papers/tsunami.pdf.

\bibitem{He02}
E.~He, J.~Leigh, O.~Yu, and T.~A. Defanti, ``{Reliable Blast UDP: predictable
  high perfromance bulk data transfer},'' in \emph{IEEE Cluster
  Computing}.\hskip 1em plus 0.5em minus 0.4em\relax IEEE Computer Society,
  2002, pp. 317--324.

\bibitem{PlanetLab}
{PlanetLab}, ``http://www.planet-lab.org/.''

\bibitem{valiant90}
L.~Valiant, ``A bridging model for parallel computation,'' \emph{Communication
  of the ACM}, vol.~33, pp. 103--111, Aug 1990.

\bibitem{Thakur05}
R.~Thakur, R.~Rabenseifner, and W.~Gropp, ``{Optimization of Collective
  Operations in MPICH},'' \emph{International Journal of High Performance
  Computing}, vol.~19, pp. 49--66, 2005.

\bibitem{Batcher68}
K.~Batcher, ``{Sorting Networks and their applications.}'' in \emph{Proceedings
  of the AFIPS Spring Joint Computing Computers}, 1968, pp. 307--314.

\bibitem{Juurlink98}
H.~A.~G. Juurlink, B. H. H.and~Wijshoff, ``{A quantitative comparison of
  parallel computation models},'' \emph{ACM Trans. Comput. Syst.}, vol.~16,
  no.~3, pp. 271--318, 1998.

\bibitem{Shroff99}
M.~Shro and R.~Geijn, ``{Collmark mpi collective communication benchmark},''
  The University of Texas at Austin,'' Technical report, December 1999.

\bibitem{Maggs95}
B.~Maggs, L.~Matheson, and R.~Tarjan, ``{Models of parallel computation: A
  survey and synthesis},'' in \emph{Proc. 28th Hawaii Int. Conf. on System
  Sciences (HICSS)}.\hskip 1em plus 0.5em minus 0.4em\relax IEEE, Jan 1995, pp.
  61--70.

\bibitem{8}
S.~Fortune and J.~Wyllie, ``Parallelism in random access machines,'' in
  \emph{STOC '78: Proceedings of the tenth annual ACM symposium on Theory of
  computing}.\hskip 1em plus 0.5em minus 0.4em\relax New York, NY, USA: ACM
  Press, 1978, pp. 114--118.

\bibitem{15}
P.~B. Gibbons, ``{A more practical PRAM model},'' in \emph{SPAA '89:
  Proceedings of the first annual ACM symposium on Parallel algorithms and
  architectures}.\hskip 1em plus 0.5em minus 0.4em\relax New York, NY, USA: ACM
  Press, 1989, pp. 158--168.

\bibitem{21}
R.~Cole and O.~Zajicek, ``{The APRAM: incorporating asynchrony into the PRAM
  model},'' in \emph{SPAA '89: Proceedings of the first annual ACM symposium on
  Parallel algorithms and architectures}.\hskip 1em plus 0.5em minus
  0.4em\relax New York, NY, USA: ACM Press, 1989, pp. 169--178.

\bibitem{12}
A.~Aggarwal, A.~Chandra, and M.~Snir, ``{Communication complexity of PRAMs},''
  \emph{Theor. Comput. Sci.}, vol.~71, no.~1, pp. 3--28, 1990.

\bibitem{16}
A.~Aggarwal, A.~K. Chandra, and M.~Snir, ``{On communication latency in PRAM
  computations},'' in \emph{SPAA '89: Proceedings of the first annual ACM
  symposium on Parallel algorithms and architectures}.\hskip 1em plus 0.5em
  minus 0.4em\relax New York, NY, USA: ACM Press, 1989, pp. 11--21.

\bibitem{24}
A.~Bar-Noy and S.~Kipnis, ``{Designing broadcasting algorithms in the postal
  model for message-passing systems},'' in \emph{SPAA '92: Proceedings of the
  fourth annual ACM symposium on Parallel algorithms and architectures}.\hskip
  1em plus 0.5em minus 0.4em\relax New York, NY, USA: ACM Press, 1992, pp.
  13--22.

\bibitem{13}
D.~Culler, R.~Karp, D.~Patterson, A.~Sahay, K.~Schauser, E.~Santos,
  R.~Subramonian, and T.~Eicken, ``{LogP: towards a realistic model of parallel
  computation},'' in \emph{{PPOPP '93: Proceedings of the fourth ACM SIGPLAN
  symposium on Principles and practice of parallel programming}}.\hskip 1em
  plus 0.5em minus 0.4em\relax New York, NY, USA: ACM Press, 1993, pp. 1--12.

\bibitem{30}
G.~Bilardi, C.~Fantozzi, A.~Pietracaprina, and G.~Pucci, ``{On the
  Effectiveness of {D--{BSP}} as a Bridging Model of Parallel Computation},''
  in \emph{ICCS '01: Proceedings of the International Conference on
  Computational Science-Part II}.\hskip 1em plus 0.5em minus 0.4em\relax
  London, UK: Springer-Verlag, 2001, pp. 579--588.

\bibitem{29}
P.~Torre and C.~Kruskal, ``{Submachine locality in the bulk synchronous
  setting.(Extended Abstract)},'' in \emph{Euro-Par '96: Proceedings of the
  Second International Euro-Par Conference on Parallel Processing-Volume II},
  vol. 1124.\hskip 1em plus 0.5em minus 0.4em\relax London, UK:
  Springer-Verlag, August 1996, pp. 352--358.

\bibitem{23}
B.~Juurlink and H.~Wijshoff, ``{The E-BSP Model: Incorporating Unbalanced
  Communication and General Locality into the BSP Model},'' \emph{In Proc.
  Euro-Par'96}, vol. 1124, pp. 339--347, January 1996.

\bibitem{26}
F.~Dehne, A.~Fabri, and A.~Rau-Chaplin, ``{Scalable Parallel Geometric
  Algorithms for Coarse Grained Multicomputers},'' \emph{In Proc. ACM 9th
  Annual Computational Geometry}, pp. 298--307, 1993.

\bibitem{27}
F.~Dehne, C.~Kenyon, and A.~Fabri, ``{Scalable Architecture Independent
  Parallel Geometric Algorithms with HIgh Probability Optimal Times},''
  \emph{In Proc. 6th IEEE Symposium on Parallel and Distributed Processing},
  pp. 586--593, Oct 1994.

\bibitem{28}
F.~Dehne, X.~Deng, P.~Dymond, A.~Fabri, and A.~Kokhar, ``{A randomized parallel
  3D convex hull algorithm for coarse grained multicomputers},'' in \emph{SPAA
  '95: Proceedings of the seventh annual ACM symposium on Parallel algorithms
  and architectures}.\hskip 1em plus 0.5em minus 0.4em\relax New York, NY, USA:
  ACM Press, 1995, pp. 27--33.

\bibitem{25}
F.~Dehne, ``{Coarse grained parallel algorithms},'' \emph{Special issue of
  Algorithmica}, vol.~24, no. 3/4, pp. 173--426, 1999.

\bibitem{37}
B.~H.~H. Juurlink and H.~A.~G. Wijshoff, ``{The Parallel Hierarchical Memory
  Model},'' in \emph{SWAT '94: Proceedings of the 4th Scandinavian Workshop on
  Algorithm Theory}.\hskip 1em plus 0.5em minus 0.4em\relax London, UK:
  Springer-Verlag, 1994, pp. 240--251.

\bibitem{38}
\BIBentryALTinterwordspacing
Z.~Li and J.~H. Mills, P. H.and~Reif, ``{Models and Resource Metrics for
  Parallel and Distributed Computation},'' in \emph{Proceedings of the
  Twenty-Eighth Annual Hawaii International Conference on System Sciences},
  Hawaii, 1995, pp. 51--60. [Online]. Available:
  \url{citeseer.ist.psu.edu/li89models.html}
\BIBentrySTDinterwordspacing

\bibitem{williams00}
\BIBentryALTinterwordspacing
T.~Williams, ``{A General-Purpose Model for Heterogeneous Computation, Ph.D.
  Thesis.}'' 2000. [Online]. Available:
  \url{citeseer.ist.psu.edu/williams00generalpurpose.html}
\BIBentrySTDinterwordspacing

\bibitem{Martin04}
J.~Martin and A.~Tiskin, ``{Dynamic BSP: Towards a Flexible Approach to
  Parallel Computing over the Grid},'' in \emph{Communicating Process
  Architectures}, I.~East, J.~Martin, and P.~Welch, Eds.\hskip 1em plus 0.5em
  minus 0.4em\relax IOS Press, 2004, pp. 219--226.

\bibitem{Vasilev03}
V.~Vasilev, ``{BSPGRID: Variable Resources Parallel Computation and
  Multiprogrammed Parallelism.}'' \emph{Parallel Processing Letters}, vol.~13,
  no.~3, pp. 329--340, 2003.

\bibitem{padyee98}
J.~Padhye, V.~Firoui, D.~Towsley, and J.~Kurose, ``{Modeling TCP throughput: a
  simple model and its empirical validation},'' in \emph{SIGCOMM '98:
  Proceedings of the ACM SIGCOMM '98 conference on Applications, technologies,
  architectures, and protocols for computer communication}.\hskip 1em plus
  0.5em minus 0.4em\relax New York, NY, USA: ACM Press, 1998, pp. 303--314.

\end{thebibliography}
\end{document}